%% file: main.tex
\documentclass[11pt,a4paper]{article}

\usepackage[T1]{fontenc}
\usepackage[utf8]{inputenc}
\usepackage[margin=1in]{geometry}
\usepackage{graphicx}
\usepackage{tikz}
\usetikzlibrary{arrows.meta}
\usepackage{amsmath,amssymb}
\usepackage{algorithm}
\usepackage{algpseudocode}
\usepackage{longtable}
\usepackage{array}
\usepackage[round,authoryear]{natbib}
\usepackage{placeins}
\usepackage{chngcntr}
\usepackage{url,lineno,microtype}
\usepackage{setspace}
\usepackage{hyperref}

\hypersetup{
  colorlinks=true,
  linkcolor=blue,
  citecolor=blue,
  urlcolor=blue
}

\counterwithout{equation}{section}

\begin{document}
\singlespacing
\InputIfFileExists{figures/stats.tex}{}{C
  \InputIfFileExists{stats.tex}{}{%
    \errmessage{stats.tex not found: upload figures/stats.tex or place stats.tex alongside main.tex}%
  }%
}
\InputIfFileExists{figures/robustness_macros.tex}{}{C
  \InputIfFileExists{robustness_macros.tex}{}{%
    \errmessage{robustness_macros.tex not found: upload figures/robustness_macros.tex or place it alongside main.tex}%
  }%
}
\def\ExpectedModelVersion{clarity-v48}
\ifdefined\StatsModelVersion
  \edef\ActualModelVersion{\StatsModelVersion}
  \ifx\ActualModelVersion\ExpectedModelVersion\else
    \errmessage{Statistics use an older model: rerun simulations and scripts/export_stats.py}
  \fi
\else
  \errmessage{Statistics predate MA detection: rerun simulations and scripts/export_stats.py}
\fi
\ifdefined\RobustnessModelVersion
  \edef\ActualRobustnessVersion{\RobustnessModelVersion}
  \ifx\ActualRobustnessVersion\ExpectedModelVersion\else
    \errmessage{Five-run macros use an older model: rerun scripts/aggregate_robustness.py}
  \fi
\else
  \errmessage{Five-run macros not loaded: rerun scripts/aggregate_robustness.py}
\fi

\title{Recognizing Mind Wandering: A Computational Phenomenology of Focused-Attention Meditation}

\author{%
P.\,C.~Kavi$^{1,*}$, D.\,A.~Friedman$^{2}$, G.~Patow$^{1,3}$\\[0.7em]
\small $^{1}$Universitat de Girona, Girona 17003, Spain\\
\small $^{2}$Active Inference Institute, CA 95531, USA\\
\small $^{3}$Universitat Pompeu Fabra, Barcelona 08002, Spain\\[0.4em]
\small $^{*}$Correspondence: Prakash Chandra Kavi, \texttt{prakash.kavi@gmail.com}%
}

\date{}

\maketitle

\begin{abstract}
Mind wandering often continues without recognition, raising a question about the relationship between attention and consciousness. How can a distraction be experienced before it is recognized as an attentional lapse, and what does recognition change? We address it for mind wandering, a form of spontaneous thought, during focused-attention meditation on the breath, through a computational phenomenology informed by Buddhist analyses of mind and meditation neuroscience. In an active-inference-inspired three-layer architecture, mental contents, termed thoughtseeds, are inferred from simulated large-scale brain network activity and compete for a capacity-limited global workspace, whose selection parallels Bengio's consciousness prior, while a separate graded monitor supports discrete recognition of lapses. Within a specified cycle of Breath Focus, Mind Wandering, Meta Awareness and Redirect Attention, recognition could begin while a distraction still held the workspace, an overlap that was usually brief. Stochastic monitoring checks also allowed intermittent recognition during ongoing wandering, with long intervals unrecognized. Recognition releases stabilization of the current regime and modulates the return policy in Meta Awareness, while return remains possible without it. Breath re-entry, recognition and redirection need not coincide. The model thus separates access to distracting content, monitoring of attention and recognition of a lapse, and shows how awareness of wandering can change its regulation without immediately ending it. This framework could be adapted to other meditation practices, to practice-related differences between novice and expert meditators, and to modeling subjective experience in other cognitive tasks.
\end{abstract}


\medskip

\section{Introduction}
Experiencing a distraction and recognizing it as an attentional lapse are distinct. An account of attentional regulation must therefore distinguish the content currently experienced, the recognition that attention has drifted from its object, and the ongoing monitoring that supports such recognition \citep{Schooler2002,Schooler2011,Dunne2019}. We use \emph{access} in Block's sense, in which content is available for reasoning, report and the guidance of action \citep{Block1995}. Content that has gained access is termed \emph{experienced content}, and a distraction that has gained access an \emph{experienced distraction}. Theories of conscious access describe how content becomes globally available for report and flexible use \citep{DehaeneChangeux2011, Mashour2020}, whereas accounts of meta-awareness address how ongoing experience itself becomes available to monitoring and control \citep{Schooler2011,Dunne2019}. These perspectives motivate a question about their interaction. How does an experienced distraction come to be recognized as a lapse, and what does recognition change?

Meditation practices differ in their cognitive aims and strategies \citep{Dahl2015} and can be described along dimensions that include attentional stability and meta-awareness \citep{Lutz2008,Lutz2015,LaukkonenSlagter2021}. Breath-directed Focused Attention (FA) practice (\emph{\=an\=ap\=anasati}) offers a direct setting for this question, because it involves sustaining attention on the breath, recognizing distraction, and repeatedly returning to the breath \citep{Lutz2008,Analayo2018}. Mind wandering is one form of spontaneous thought \citep{Christoff2016}, and in FA practice it arises against an explicit intention to attend to the breath.

We adopt the four-phase canonical cycle of FA practice (Fig.~\ref{fig_1}) as a working description \citep{Hasenkamp2012,Lutz2008}. Two things change from phase to phase. One is what attention is on. The other is whether the lapse, the departure of attention from its object, is known. In this canonical sequence, the lapse becomes known in Meta Awareness (MA), before attention returns to the breath during Redirect Attention. Knowing the lapse and returning to the breath are therefore distinct. Mind wandering with and without meta-awareness recruits overlapping neural systems to different degrees, supporting the distinction between distracting content and awareness of that content \citep{Christoff2009}. The aim of practice is sustained awareness of the natural breath \citep{Hart1988,Analayo2018}. Early in training this requires repeated recognition and deliberate return. As practice develops, these regulative skills are needed less often, and sustaining focus is described as becoming progressively effortless \citep{Lutz2008}.

\begin{figure}[tbp]
    \centering
    \input{figures/fig1-tikz.tex}
    \caption{\textbf{The Canonical Cycle of Focused-Attention Meditation.} The cycle is conceptualized as four distinct phases, following \citet{Hasenkamp2012}. (1) Breath Focus (BF) is sustained attention on the breath \citep{Lutz2008,MacLean2010}, associated with dorsolateral prefrontal activity \citep{Hasenkamp2012}. (2) In Mind Wandering (MW), attention drifts from the breath to thoughts or other distractions, with the lapse not yet known in this canonical sequence, associated with default mode network (DMN) activity \citep{Christoff2016,Fox2015}. (3) Meta Awareness (MA) is the interval in which the lapse can be recognized, engaging salience-network regions \citep{Hasenkamp2012,Schooler2011,Seeley2007}. (4) Redirect Attention (RA) disengages from distracting content and returns attention to the breath, engaging frontoparietal executive regions \citep{Hasenkamp2012,Lutz2008}. Breathing continues in every phase. The cycle is a canonical idealization. In practice the order can vary, for example when the mind wanders again before attention is redirected.}
    \label{fig_1}
    \par\smallskip\noindent{\small Alt text: Four labeled boxes arranged in a loop, Breath Focus, Mind Wandering, Meta Awareness and Redirect Attention, joined by curved arrows in the canonical order of the focused-attention cycle, with a short label on each arrow naming the transition.}
\end{figure}

Relating such first-person distinctions to neural dynamics calls for explicit links to computational mechanisms \citep{Lutz2025}. Computational phenomenology provides such links by relating descriptions of lived experience to models of the processes that organize them \citep{Ramstead2022,SandvedSmith2021,SandvedSmith2025}, extending Varela's program of bringing cognitive science and human experience together \citep{Varela2017}. In the model, each phase becomes a discrete \emph{regime}. The four-regime organization is specified, while access, monitoring and control shape the realized sequence. We treat expertise as graded. The novice and the expert pass through the same cycle, with different dwell times, transition tables and network attractors.

Predictive-processing accounts have interpreted focused-attention meditation in terms of precision weighting and the regulation of habitual responses, motivating explicit computational simulations \citep{Lutz2019}. Existing computational models address complementary aspects of lapse recognition and attentional regulation. A deep parametric active inference model of attentional control and meta-awareness uses nested inference over cognitive states and their precision, and simulates focus, unrecognized distraction, recognition and refocusing \citep{SandvedSmith2021}. The predictive global neuronal workspace connects active inference with ignition and report in visual consciousness \citep{Whyte2021}. Recurrent-network models have also examined autonomous shifts between focus and mind wandering through prediction-error-based regulation of predictive processing \citep{Oyama2025}, and meta-attentional control of present-centered interoception and internally simulated activity within an allostatic framework \citep{Idei2026}. Neither of these recurrent-network models represents the recognition of a lapse as a separate event. Models developed for focused-attention meditation separate attentional control from stochastic metacognitive sampling \citep{ChristianAgrawal2022}, or model competition among task-related and mind-wandering processes, with a meta-task memory supporting a return to focus \citep{MoyeVanVugt2021}. We examine a complementary formulation in which identifiable mental contents compete for a capacity-limited workspace, while a separate graded monitor, termed \emph{monitoring clarity}, is driven by the inferred thoughtseed activations. This formulation distinguishes content access, ongoing monitoring and detection, so that their interaction can be examined within a specified focused-attention cycle. The question posed above then divides into three parts, (i) how mental content is represented, (ii) how it gains access, and (iii) how recognition arises and what it changes.

\textbf{Representing content.} Mental content is represented as \emph{thoughtseeds}, latent causes of simulated large-scale network activity. The construct draws on the Neuronal Packet Hypothesis, which proposes that neuronal packets, self-organizing ensembles of neurons, can form superordinate ensembles supporting cognitive functions \citep{Yufik2019,Ramstead2021}. \citet{Yufik2026} have since written that ``neuronal packets are `molecules of meaning' (or `thoughtseeds' [\citealp{Kavi2024,Kavi2025}]), coordinated packet complexes integrate non-contiguous groupings and thus impute meaning to indefinite stretches in the stimuli streams (unrelated bundles can have values but no meaning).'' How such ensembles map onto large-scale brain dynamics remains open, and the model treats thoughtseeds as latent causes without committing to a neural implementation. Most thoughtseeds are \emph{first-order} contents, what attention can be on, and these compete for workspace access. Others are \emph{monitoring} thoughtseeds, which do not enter the workspace and instead contribute to monitoring. The model implements one, related to recognizing a lapse. Conceptually, thoughtseeds form a larger \emph{dormant inventory} of which a small, context-relevant subset is active at any time. The seed metaphor also draws on Yog\=ac\=ara accounts of \emph{\=alaya-vij\~n\=ana} (storehouse consciousness), in which mental seeds (\emph{b\={\i}ja}) condition subsequent experience \citep{Waldron2003}, and the emphasis on explicitly specified conditions offers a descriptive analogy with dependent arising (\emph{prat\={\i}tyasamutp\={a}da}) \citep{Analayo2021}. Both are conceptual correspondences, and the four-phase transition structure is specified independently of either. Supplementary Section~S5 details the contemplative motivations and their limits.

\textbf{Gaining access.} Content representation is distinct from access to that content. Global Neuronal Workspace (GNW) theory emphasizes the amplification and broadcast that make selected content available for conscious access \citep{Mashour2020,DehaeneChangeux2011}. Selection into the workspace parallels Bengio's consciousness prior, in which attention extracts a low-dimensional conscious state from a high-dimensional representation \citep{Bengio2017}, and limits on immediate memory offer a behavioral parallel for capacity constraints \citep{Miller1956}. Abhidhamma analyses, in which each moment of consciousness (\emph{citta}) takes a single object, offer a contemplative parallel for single-content access \citep{Bodhi2000}. In the model, a capacity-limited workspace selects among the active thoughtseeds. Access is content-specific, and accessed content supplies evidence for lapse recognition and contributes to control. The model addresses functional access and meta-awareness, not phenomenal consciousness.

\textbf{Recognizing a lapse.} Workspace access addresses the selection problem but leaves open the regulation problem, because an experienced distraction need not be recognized as an attentional lapse. Recognizing a lapse depends on monitoring, which is distinct from the attention regulation it supports \citep{Lutz2008}. Meta-awareness can be sustained and need not be propositional \citep{Dunne2019,Schooler2011}. In the canonical cycle, awareness of wandering precedes redirection \citep{Hasenkamp2012}. Yet several alternations between focus and off-task thought can occur before a lapse is noticed \citep{ZukoskyWang2021}. Distracting content that holds attention can also sustain wandering, consistent with the persistence of task-unrelated thought \citep{Levinson2012}. The dynamic framework of spontaneous thought describes such persistence in terms of automatic constraints, which hold attention on salient content outside cognitive control, as distinct from deliberate constraints exercised through it \citep{Christoff2016}. Computational accounts of meta-awareness and self-monitoring motivate distinguishing monitoring from the content it monitors \citep{SandvedSmith2021,DehaeneLauKouider2017,Fleming2020}. In the model, the experienced content represents what attention is on, while a separate belief tracks whether sustained breath focus is established. \emph{Detection} marks the onset of a known lapse, and the lapse stays known while that belief and monitoring support it. A known lapse can shape subsequent regulation without immediately displacing the distraction or guaranteeing return to the breath (Section~\ref{sec:recognition}).

\begin{figure}[tbp]
  \centering
  \input{figures/fig2-tikz.tex}
  \caption{\textbf{Conceptual architecture.}
  An illustrative representation of a meditating agent, focusing on its internal states. \textbf{Layer 1 (High-dimensional brain networks)} is the neuronal substrate, simulated as network activation profiles. Superordinate ensembles of neuronal packets (dashed) motivate the hierarchy \citep{Yufik2019,Ramstead2021,Hipolito2021,Yufik2026}. Layer 1 also holds the configured regime transitions, a dwell-time distribution for each regime and a transition table. \textbf{Layer 2 (Low-dimensional latent causes)} holds the thoughtseed repertoire, of which the context-active thoughtseeds are a small subset, and relays the control formed in Layer 3 to Layer 1. Circles mark the contents that can enter the workspace, with the accessed one in orange, and the diamond marks the recognition cue, which does not enter it. \textbf{Layer 3 (Metacognition)} contains the global workspace \citep{DehaeneChangeux2011,Mashour2020} with its learned access prior. Capacity-limited access (funnel) straddles the L2--L3 boundary and passes candidate contents from L2 to the workspace. The accessed content is the evidence for the lapse belief and detection, and it is forwarded for regulation. Monitoring clarity (graded) rises and falls with the inferred thoughtseed activations, whereas detection is a discrete event that marks the onset of an interval in which the lapse is known. Ascending evidence (red, perception) and descending control (blue, action) couple the layers \citep{FristonParrDeVries2017}. The Markov blanket of the agent (sensory and active states) and the external states (environment) are shown for context. The nested outlines indicate conceptual organization, not established conditional-independence boundaries. Solid component boxes are simulated and dashed boxes give conceptual context. The implemented model is shown in Fig.~\ref{fig_3}.}
  \label{fig_2}
  \par\smallskip\noindent{\small Alt text: Schematic of a meditating agent as three nested layers, with brain networks at the bottom, a thoughtseed repertoire in the middle, and a global workspace with monitoring, lapse belief, detection and return policy at the top. Red arrows carry ascending evidence and blue arrows descending control.}
\end{figure}

The motivations above converge in a three-layer architecture (Fig.~\ref{fig_2}). L1 is a physiological generative process, L2 infers thoughtseeds as latent causes of network activity, and L3 is a capacity-limited workspace with monitoring and control. L1 simulates large-scale brain networks because research on meditation and mind wandering has linked attentional regulation to interactions among them \citep{BrandmeyerDelorme2021,Tang2015,Fox2016}. The Free Energy Principle and active inference provide a framework for the inference and control in the architecture by relating perception, learning, and action to generative models and variational free-energy minimization \citep{Friston2010,Friston2017,Ramstead2020}. This connects inference about mental content with regulation of the physiological activity that supports it. The simulation is a restricted, active-inference-inspired implementation of this framework (Sections~\ref{sec:architecture} and~\ref{sec:recognition}), and it abstracts from exteroceptive input to examine an internal physiological generative process, motivated by interoceptive predictive accounts \citep{BarrettSimmons2015}. In paired novice and expert simulations, we examine how content access, monitoring and recognition interact to shape distraction and return to the breath.

\FloatBarrier
\section{Materials and Methods}

\subsection{Model Scope, Generative Process and Simulation Design}\label{sec:scope}

The simulation takes the four-phase cycle of Fig.~\ref{fig_1} as a specification, with each phase a regime $s_t$ (BF, MW, MA or RA). Regimes are mutually exclusive, while network and thoughtseed activations are continuous. MA is a configured interval that favors conditions for recognition, and L1's transition process determines when it is entered. Recognition itself is the L3 detection event. It does not trigger entry into MA and can also occur outside MA (Section~\ref{sec:recognition}). Novice and expert phenotypes run the same model. They differ only in their dwell-time distributions, transition tables and network attractors, and share all other settings, including the thoughtseed priors and the learning rate.

L1 is the generative process for L2 and L3, the simulated physiological dynamics that they infer and regulate through interfaces between adjacent layers, modeled on Markov blankets \citep{Kirchhoff2018} (Fig.~\ref{fig_3}). This follows hierarchical accounts in which each level treats the level below as its generative process \citep{Hipolito2021}. The regime is a controlled explicit-duration semi-Markov process \citep{Yu2010}. A phenotype- and regime-specific Gamma dwell-time distribution with mean $T_s$ gives a baseline exit probability $h_s(a)$ that depends on the time already spent in the regime, and on exit the successor is drawn from the phenotype's transition table. Broadcast stabilization lowers this probability, and in MA the return policy also modulates exit timing and supplies the successor (Section~\ref{sec:recognition}).

A switching multivariate Ornstein--Uhlenbeck (OU) process evolves seven network activations $\mathbf{x}_t$ \citep{Gilson2016,Yeo2011}, namely default mode (DMN), ventral attention (VAN), dorsal attention (DAN), frontoparietal (FPN), visual (VIS), somatomotor (SOM), and limbic (LIM). VAN includes the anterior insular and dorsal anterior cingulate territory associated with salience processing \citep{Seeley2007}. Activity relaxes toward a regime-specific attractor \citep{Friston2014},
\begin{equation}
d\mathbf{x}_t = -\Theta(s_t)\bigl(\mathbf{x}_t-\boldsymbol{\mu}_x(s_t)\bigr)\,dt + \sigma_x\, d\mathbf{W}_t,
\label{eq:l1}
\end{equation}
where $\Theta(s_t)$ is a regime-specific coupling matrix, stable by construction, and $\boldsymbol{\mu}_x(s_t)$ is the attractor of the current regime. The attractors are configured profiles rather than measured signatures. Default-mode activity is not a unique marker of mind wandering \citep{Kucyi2016,Groot2021}. Supplementary Section~S1.2 gives the coupling matrices, the discretization and the duration model.

The model implements five context-active thoughtseeds \citep{Kavi2025,Kavi2026}, named for experiences described in accounts of breath-directed FA practice \citep{Hart1988,Analayo2018}. Four are first-order contents, what attention can be on, namely the meditation object \textit{breath\_sensation}, the distractors \textit{pain\_discomfort} and \textit{pending\_tasks}, and \textit{intention\_to\_return}. The two distractors represent bodily discomfort and self-generated thought, which the Mind Wandering regime groups together despite their distinct phenomenological character. The fifth, \textit{recognition\_cue}, is a monitoring thoughtseed, a graded L2 activation. It adds a separate term to the drive of monitoring clarity and does not hold the workspace. Each thoughtseed has a configured prior mean activation in every regime, collected in $\boldsymbol{\mu}_z(s)$. MA has the priors of MW for the four first-order contents, which allows distracting content to remain active in MA, and only \textit{recognition\_cue} differs, with its highest prior in MA. Supplementary Section~S5 gives fuller descriptions and their contemplative correspondences.

Table~\ref{tab:notation} lists the main symbols by layer and kind. Agreement with configured quantities verifies the implementation, whereas emergent quantities such as workspace access, monitoring clarity and detection are the outcomes the model examines (Section~\ref{sec:empirical}). Supplementary Table~S6 gives the phenomenological interpretation of each quantity and its limit.

{\small
\setlength{\LTpre}{6pt}%
\setlength{\LTpost}{6pt}%
\setlength{\LTcapwidth}{\textwidth}%
\begin{longtable}{p{2.4cm} p{0.62\linewidth} >{\raggedright\arraybackslash}p{2.0cm}}
\caption{\textbf{Symbol glossary.} Main symbols of Sections~\ref{sec:scope} to~\ref{sec:recognition}, Fig.~\ref{fig_3} and Algorithm~\ref{alg:step}. The third column gives the kind of each quantity. Configured and derived quantities are fixed (grey in Fig.~\ref{fig_3}), and derived ones are computed once from configured settings. Learned quantities, and those computed from learned counts, change during training only (yellow). Computed quantities are evaluated during the simulation (white). Supplementary Section~S2 gives the values.}
\label{tab:notation}\\
\multicolumn{3}{l}{\textbf{Layer 1: Physiological Generative Process}} \\
\hline
$s_t$ & Regime (BF, MW, MA or RA) & computed \\
$\mathbf{x}_t$ & Network activations (DMN, VAN, DAN, FPN, VIS, SOM, LIM) & computed \\
$\boldsymbol{\mu}_x(s)$ & Network attractor for regime $s$ & configured \\
$T_s$ & Mean dwell time of regime $s$ & configured \\
$h_s(a)$ & Baseline exit probability of regime $s$ on its $a$-th step, from the Gamma dwell-time distribution with mean $T_s$ & derived \\
$S_t$ & Broadcast stabilization & computed \\
$\tilde h_t$ & Stabilized exit probability & computed \\
\\
\multicolumn{3}{l}{\textbf{Layer 2: Thoughtseed Inference and Control}} \\
\hline
$\mathbf{z}_t^{*}$ & Corrected activations of the five thoughtseeds, passed to L3 & computed \\
$\rho_t(s)$ & L2's belief over the regime & computed \\
$W,\mathbf{b}$ & Linear decoder from thoughtseed activations to network activity & learned \\
$\Pi_x$ & Observation precision & learned \\
\\
\multicolumn{3}{l}{\textbf{Layer 3: Workspace Access, Monitoring, Detection and Return Policy}} \\
\hline
$o_t$ & Accessed thoughtseed, one of the four first-order thoughtseeds, or empty slot $\varnothing$ & computed \\
$\boldsymbol{\beta}$ & Access prior over the four first-order thoughtseeds & learned \\
$m_t$ & Monitoring clarity & computed \\
$m_{\mathrm{rest}}$ & Resting monitoring clarity, the target of $m_t$ at zero contrast & from learned counts \\
$e_t$ & Check-interval indicator, 1 for $\tau_m$ from a monitoring check & computed \\
$q_t$ & Lapse belief, the estimated probability that sustained breath focus is not established & computed \\
$R_t$ & Departure-evidence flag, set when a distractor holds the workspace & computed \\
$\psi(m,q)$ & Detection condition, $m>m_{\mathrm{rest}}$ and $q\ge\tfrac12$ & computed \\
$D_t,\Delta_t$ & Recognition indicator, and the detection event at its onset (recognition onset) & computed \\
$E$ & Return habit over the MA successors, returning (RA) and not returning (MW) & learned \\
$\gamma_t$, $P(\pi)$, $\Lambda_t$ & Policy precision, return policy and MA exit-odds factor & computed \\
\\
\multicolumn{3}{l}{\textbf{Shared Configuration and Derived Quantities}} \\
\hline
$\boldsymbol{\mu}_z(s)$ & Thoughtseed prior means for regime $s$ & configured \\
$d_i(s)$ & Diagnosticity, the share of the prior of thoughtseed $i$ that falls in regime $s$ & derived \\
$c(i)$, $w_i$ & On-task contrast, the log ratio of the mean prior of $i$ outside MW to its prior in MW, and on-task weight $w_i=\tanh[c(i)/2]$ & derived \\
$\lambda(o)$ & Lapse evidence, the log ratio of the mean prior of $o$ over MW, MA and RA to its prior in BF, with $\lambda(\varnothing)=0$ & derived \\
$p_k$ & Probability of a monitoring check per step while a distractor holds the workspace & configured \\
$c_k$ & Check contrast, the \textit{recognition\_cue} term of $\bar c_t$ at the cue's MA prior & derived \\
$G(\pi)$ & Policy score of the regime entered by $\pi$ & derived \\
\end{longtable}
}

Each phenotype is trained for \NovicePracticeSteps{} steps and then runs a \StatsEvaluationSteps{}-step frozen rollout, which keeps the learned quantities and resets transient states. Figures and outcome measures use the evaluation window, the final \StatsPlotSteps{} rollout steps. Five paired runs with different random seeds provide a qualitative robustness check. Specification, parameters and outcome definitions are in Supplementary Sections~S1--S3.

\subsection{Inference, Access and Monitoring}\label{sec:architecture}
L2 and L3 infer and regulate the L1 process of Section~\ref{sec:scope}. Each timestep of $dt=0.2$\,s runs the recurrent loop of Fig.~\ref{fig_3} in the order of Algorithm~\ref{alg:step}. No layer receives the regime from another. L2 infers it from the network activity, and L3 receives only the corrected thoughtseed activations (Supplementary Section~S1.5). Step 1 is the L1 update of Section~\ref{sec:scope}, which applies the descending control of step 5 of the previous timestep. The paragraphs below and in Section~\ref{sec:recognition} follow steps 2 to 5.

\begin{figure}[tbp]
  \centering
  \input{figures/fig3-tikz.tex}
  \caption{\textbf{Implemented model and update order.} Conventions follow Fig.~\ref{fig_2}, with ascending evidence in red, descending control in blue, the accessed thoughtseed as an orange dot, the recognition cue, which does not hold the workspace, as a green diamond, and lapse belief and detection as a tint. L2 is not given the regime and infers it from the network activity, scored with the learned decoder and observation precision. Per step, L3 receives only the corrected thoughtseed activations, and the quantities derived from the configured content priors (grey arrow) are fixed. Circled numbers are the lines of Algorithm~\ref{alg:step}, which gives the order and timing within one step. Training-only updates are not drawn. Grey boxes are configured, yellow boxes are learned, and white boxes are computed during the simulation. The realized regime and L2's regime belief are computed, so their boxes are white. Their grey insets are the configured dwell-time distributions and transition table, and the transition model that L2 builds from them. The layers of Fig.~\ref{fig_2} are drawn as stacked lanes rather than nested shells, to show the signals exchanged between adjacent layers and their order within one step.}
  \label{fig_3}
  \par\smallskip\noindent{\small Alt text: Three horizontal lanes for Layers 1 to 3 show the signals exchanged in one timestep. Network activity rises from Layer 1 to Layer 2, which infers the regime from it, corrected thoughtseed activations rise to the Layer 3 workspace, which holds the learned access prior, and the accessed content, the recognition indicator and the return policy's successor probabilities and exit-odds factor descend to Layer 1, which applies them at the next step and forms the broadcast stabilization. Circled numbers 1 to 5 mark the update order. In the thoughtseed box, four circles are the workspace candidates, one orange for the accessed content, and a diamond set apart marks the recognition cue. Grey boxes are configured, yellow boxes learned and white boxes computed.}
\end{figure}

\begin{algorithm}[tbp]
\caption{\textbf{One simulation step.} Lines 1--5 carry the circled numbers of Fig.~\ref{fig_3}, and the unnumbered final line runs in training only. The quantities forwarded in line 5 are applied in line 1 of the next step.}
\label{alg:step}
\begin{algorithmic}[1]
\State \textbf{Network dynamics and regime (L1).} Test for a regime exit with the stabilized exit probability $\tilde h_{t-1}$ (Eq.~\eqref{eq:act}), whose odds are multiplied in MA by $\Lambda_{t-1}$ (Supplementary Section~S1.4). On exit, draw the successor from the transition table, or in MA from the return policy formed at $t-1$. Advance $\mathbf{x}_t$ toward the attractor of $s_t$ (Eq.~\eqref{eq:l1})
\State \textbf{Content inference (L2).} Predict the regime belief and each regime's thoughtseed activations, update $\rho_t$ from $\mathbf{x}_t$, correct each prediction and combine the corrections into $\mathbf{z}_t^{*}$, which passes to L3 (Supplementary Section~S1.3)
\State \textbf{Access (L3).} Select $o_t$ from $\mathbf{z}_t^{*}$, $\beta$ and $m_{t-1}$ (Eq.~\eqref{eq:access})
\State \textbf{Monitoring, lapse belief, detection and return policy (L3).} Draw a monitoring check if $o_t$ is a distractor, and update $m_t$ (Eq.~\eqref{eq:gate}). Update $q_t$ from $o_t$ at precision $m_{t-1}$ (Eq.~\eqref{eq:belief}) and then $R_t$. Evaluate $D_t$ from $\psi(m_t,q_t)$, $R_t$ and $D_{t-1}$, and its onset $\Delta_t$ (Eq.~\eqref{eq:detect}). Form $P(\pi)$ and $\Lambda_t$ with $\gamma_t=D_t\,m_tq_t$ (Eq.~\eqref{eq:policy})
\State \textbf{Descending control (L3$\to$L2$\to$L1).} L2 sends $o_t$, $D_t$, $P(\pi)$ and $\Lambda_t$ to L1, which forms $S_t$ from $o_t$ and $D_t$ (Eq.~\eqref{eq:act}) at its next update
\Statex \textbf{Learning (L2 and L3, training only).} Update the decoder and $\Pi_x$, the access prior $\boldsymbol{\beta}$ and the access counts, and if $\Delta_t=1$ add $P(\pi)$ to the counts of the habit $E$ (Supplementary Section~S1.6)
\end{algorithmic}
\end{algorithm}

\textbf{Step 2, content inference (L2).} L2 updates its regime belief $\rho_t(s)$ and the thoughtseed activations $\mathbf{z}_t^{*}$ with a switching Kalman filter \citep{GhahramaniHinton2000}. The belief is predicted with a transition model built from the phenotype's configured dwell means and transition table. Each regime predicts the thoughtseed activations from the preceding estimate and its configured content priors $\boldsymbol{\mu}_z(s)$. The decoder maps each prediction to expected network activity. Its mismatch with $\mathbf{x}_t$, weighted by the observation precision $\Pi_x$, updates the regime belief and corrects the predictions, which are combined into $\mathbf{z}_t^{*}$ (Supplementary Section~S1.3). The decoder starts from configured network signatures, the affine map that sends each regime's thoughtseed prior means to its network attractor, and is then learned. The corrected activations stay close to the priors of the inferred regime, while the network evidence can still change access.

\textbf{Step 3, access (L3).} L3 selects which content is experienced. Its single-slot workspace holds at most one of the four first-order thoughtseeds, and \textit{recognition\_cue} is not a candidate. Each candidate $i$ has an access drive
\begin{equation}
u_{i,t}=z^*_{i,t}+\beta_i+(m_{t-1}-m_{\mathrm{rest}})\,w_i+\xi_{i,t}.
\label{eq:access}
\end{equation}
The drive adds to the inferred activation the learned access prior, a monitoring bias and an access-history term. The on-task weight $w_i$ is positive for on-task content and negative for the two distractors, so monitoring clarity above its resting value biases the competition toward on-task content \citep{DesimoneDuncan1995}. The history term $\xi_{i,t}$ favors the current content through reverberation and progressively weakens it through adaptation, as in competition models of perceptual bistability \citep{Shpiro2009}. The candidate with the highest drive becomes $o_t$ if its drive reaches the access threshold, and otherwise the slot is empty. Thresholding and reverberation give GNW-inspired all-or-none access \citep{DehaeneKerszbergChangeux1998,DehaeneChangeux2011,Mashour2020}, and sustained access is repeated selection of the same content. Supplementary Section~S1.4 gives the history term and the adaptation update, and Supplementary Table~S1 the parameter values.

\textbf{Step 4, monitoring (L3).} Monitoring clarity $m_t$ is an operational, task-aligned signal. It integrates, with a delay, how strongly the inferred activations favor on-task content, so it does not itself measure whether a lapse is recognized. It relaxes toward a bounded target,
\begin{equation}
m_t = m_{t-1} + \frac{dt}{\tau_m}\bigl(\sigma[\bar c_t+c_k e_t+\operatorname{logit} m_{\mathrm{rest}}] - m_{t-1}\bigr),
\label{eq:gate}
\end{equation}
where $\sigma$ is logistic and $\tau_m=2$\,s. The contrast $\bar c_t$ has two parts. The first is the activation-weighted mean of the on-task contrast $c(i)$ over the four first-order thoughtseeds. The second is a separate term for \textit{recognition\_cue}, proportional to its activation above its MW prior (Supplementary Section~S1.4). At zero contrast the target is the resting clarity $m_{\mathrm{rest}}$, which is also the reference for the access bias, detection and the access credit. The term $c_k e_t$ lets monitoring re-engage during ongoing distraction, not only through the \textit{recognition\_cue} term in MA. A monitoring check is a stochastic, unscheduled event. It can occur at any step at which a distractor holds the workspace, with a configured probability $p_k$ per step shared by both phenotypes, and it does not depend on the regime, monitoring clarity or the lapse belief. The model specifies when checks are eligible and their probability, but not why a particular check occurs. Each check sets $e_t$ to 1 for one monitor time constant (Supplementary Section~S1.4). A check can affect access from the next step, through monitoring clarity.

\textbf{Step 4, lapse belief (L3).} Access to a distractor does not by itself amount to recognizing the lapse \citep{Schooler2011,Levinson2012}, so L3 holds a belief $q_t$ about the attentional state, adapted from the attentional level of \citet{SandvedSmith2021}. Here $q_t$ is the estimated probability that sustained breath focus is not established, with the established state approximated by BF and the other by MW, MA and RA. Each step, the predicted belief is updated in log-odds by the lapse evidence of the accessed content,
\begin{equation}
\operatorname{logit} q_t=\operatorname{logit} q_t^{-}+m_{t-1}\,\frac{dt}{\tau_{\mathrm{dyn}}}\,\lambda(o_t),
\label{eq:belief}
\end{equation}
where $q_t^{-}$ is the belief predicted from $q_{t-1}$ with dwell-based approximations of the per-step probabilities that sustained breath focus ends and that it is re-established (Supplementary Section~S1.4), and $\tau_{\mathrm{dyn}}=1$\,s is the time constant of the thoughtseed dynamics. The lapse evidence $\lambda(o)$ is positive for the two distractors and negative for the breath sensation and, more weakly, for the intention to return. Monitoring clarity sets the precision of this evidence.

\subsection{Recognition, Return and Outcome Measures}\label{sec:recognition}
Recognition and the return policy complete step 4 of Algorithm~\ref{alg:step}, and step 5 carries them to L1 as descending control. Each is followed by its outcome measures. Recognition has an onset, marked by the detection event, and an interval during which the lapse remains recognized. Return is measured at two levels, breath re-entry into the workspace and the regime exit from MA to RA. Each run gives one summary per measure over the evaluation window, and Table~\ref{tab:event_measures} reports their median and range across the five runs. Supplementary Section~S3 gives eligibility and the treatment of episodes cut by the window, and Supplementary Table~S5 the event counts.

\textbf{Step 4, detection (L3).} Detection operationalizes recognition of the lapse. This operationalization is a mechanistic proposal, and its correspondence with subjective recognition is not assumed (Section~\ref{sec:limitations}). It uses the condition $\psi(m,q)=\mathbb{1}[m>m_{\mathrm{rest}}]\,\mathbb{1}[q\ge\tfrac12]$ and a departure-evidence flag $R_t$, set when a distractor holds the workspace and cleared when the belief is below $\tfrac12$ without a distractor present. The binary recognition indicator $D_t$ represents whether a lapse remains recognized in the model. Both $D_t$ and its onset, the detection event $\Delta_t$, follow
\begin{equation}
D_t=\psi(m_t,q_t)\,\max(D_{t-1},R_t),\qquad \Delta_t=D_t\,(1-D_{t-1}).
\label{eq:detect}
\end{equation}
The lapse belief alone therefore cannot start detection. Onset also requires the departure-evidence flag, which only a distractor in the workspace sets. The event marks recognition onset. The recognition interval follows while the indicator remains 1, and it ends at recognition offset, the first later step with $D_t=0$, when clarity has reached its resting level or the belief has fallen below $\tfrac12$. The indicator controls policy modulation and the release of stabilization, while its onset triggers habit learning. Supplementary Section~S1.4 gives the update of $R_t$. Detection does not displace the accessed content. In MA, the \textit{recognition\_cue} term of $\bar c_t$ can support monitoring clarity while distractor evidence supports a high lapse belief, allowing recognition to begin while a distractor still holds the workspace. During Mind Wandering, the check term of Eq.~\eqref{eq:gate} can provide such support. The rule permits this overlap by construction, while its frequency and duration are outcomes.

\textbf{Recognition measures.} Each detection event in the window is followed until the window ends. Breath re-entry is the first later step at which the breath sensation regains the workspace. The breath re-entry proportion is the share of all events, including those with the breath already present, followed by breath re-entry before or at recognition offset, and the recognition-to-breath latency is the time from the event to that re-entry. Same-distractor overlap runs from an event with a distractor in the workspace until recognition offset or until that distractor leaves. Recognition during established Mind Wandering counts detection events at least 3\,s after MW entry, which excludes the early onsets supported by clarity carried over from the preceding regime. It is reported per minute of such time and, over complete MW episodes longer than 3\,s within the window, as the share of episodes without a recognition onset after that point and the longest stretch of each episode without recognition.

\textbf{Step 4, return policy (L3).} For MA exits, L3 scores the two exits, returning ($\pi=\mathrm{RA}$) and not returning ($\pi=\mathrm{MW}$), against a learned habit $E$ with precision $\gamma_t=D_t\,m_tq_t$, monitoring clarity weighted by the lapse belief while the lapse is recognized,
\begin{equation}
P(\pi)=\operatorname{softmax}_\pi\bigl(\ln E(\pi)-\gamma_t\,G(\pi)\bigr).
\label{eq:policy}
\end{equation}
Here $G(\pi)$ is a one-step, active-inference-inspired score of the regime $s_\pi$ entered by $\pi$ \citep{Friston2017}. It is the cross-entropy of a proxy content forecast built from the entered regime's configured priors against a preference for contents diagnostic of BF (Supplementary Section~S1.4). Eq.~\eqref{eq:policy} has the standard form of a prior over policies, in which the precision $\gamma_t$ weights the score $G$ relative to the habit \citep{Smith2022}. Here it is computed at each step, without a separate precision update. The score is lower for returning, so as $\gamma_t$ rises the choice shifts from the habit toward returning. At $\gamma_t=0$, including whenever $D_t=0$, the policy equals the learned habit $E$, consistent with habitual policy selection at low precision \citep{Hesp2021}. The policy selects the successor regime and does not hold the workspace, and the experienced intention to return is the separate content \textit{intention\_to\_return}.

\textbf{Step 5, descending control (L3$\to$L2$\to$L1).} Layers exchange signals only with adjacent layers, so L3 influences L1 through L2. L2 forwards the accessed content, the recognition indicator, the return policy and the exit-odds factor $\Lambda_t$. At its next update, L1 uses them to set stabilization, exit timing and the successor regime. From the content and the indicator it forms the \emph{broadcast stabilization} $S_t$, which lowers the exit probability,
\begin{equation}
S_t=(1-D_t)\,d_{o_t}(s_t),\qquad
\tilde h_t=h_{s_t}(a_t+1)\,(1-S_t),
\label{eq:act}
\end{equation}
where $a_t$ is the number of steps already spent in the regime and $d_\varnothing\equiv0$ for an empty slot. First-order content holding the workspace stabilizes the current regime in proportion to its diagnosticity for that regime, so a distractor holding the workspace in MW stabilizes MW. An empty workspace provides no stabilization. Recognition releases stabilization without forcing redirection. In MA, $\Lambda_t$ scales the odds of $\tilde h_t$, using the same scores as the successor policy. Supplementary Section~S1.4 gives the resulting exit probability and how it changes with precision.

\textbf{Return measures.} Return effort is read out at an MA exit at step $t$ as $D_{\mathrm{KL}}[P_{t-1}(\pi)\,\|\,E]$, the divergence of the applied policy, formed at $t-1$, from the habit \citep{ParrHolmes2023}. It does not act on the dynamics, and its window value averages all MA exits, with or without a detection event. An MA episode that begins in the window and contains a detection event is qualifying. Its recognition-to-RA latency runs from its first detection event to RA entry when it exits directly to RA, and the within-run summary is the median, as for the recognition-to-breath latency.

\textbf{Learning.} Five quantities are learned during training, namely the decoder, the observation precision, the access prior, the access counts and the return habit. The decoder takes one gradient step per timestep on the reconstruction error, and the observation precision follows a variational-Laplace update \citep{Friston2007Laplace}. The other three are counts in Dirichlet priors, as in active-inference learning \citep{Friston2017}. The breath-access prior gains an access credit based on monitoring clarity above its resting level at re-entry, awarded after the required uninterrupted hold, so practice raises it through returns supported by monitoring, a proposed mechanism of practice learning, with its contemplative motivation in Supplementary Section~S5. The access counts record which content holds the workspace and set resting clarity. The return habit starts from the phenotype's MA row of the transition table and adds the return act formed at each detection event \citep{Friston2016Learning}. The frozen rollout omits these updates (Supplementary Section~S1.6).

\section{Simulation Results}

Figures~\ref{fig_4}--\ref{fig_6} show one illustrative run per phenotype, training seed \StatsSeed{}, chosen by inspection of the five runs as the one whose window shows the lapse, recognition and return sequence of both phenotypes most clearly. It is not a representative run, and its expert-novice differences need not follow every five-run median direction. Unless otherwise stated, summaries use the evaluation window (Section~\ref{sec:scope}), and Fig.~\ref{fig_6} shows a shorter interval around selected events. Table~\ref{tab:event_measures} summarizes recognition and return across the five runs, and Supplementary Table~S4 gives paired expert-minus-novice differences for these and other measures. These five-run summaries are the context for phenotype comparisons, and they describe stochastic variability, not population effects. Supplementary Fig.~S1 gives training and rollout diagnostics.

\subsection{Configured Dynamics, Workspace Access and Monitoring}\label{sec:verify}

\begin{figure*}[tbp]
\centering
\begin{tabular}{c c}
\small{(A)} & \includegraphics[width=0.55\textwidth]{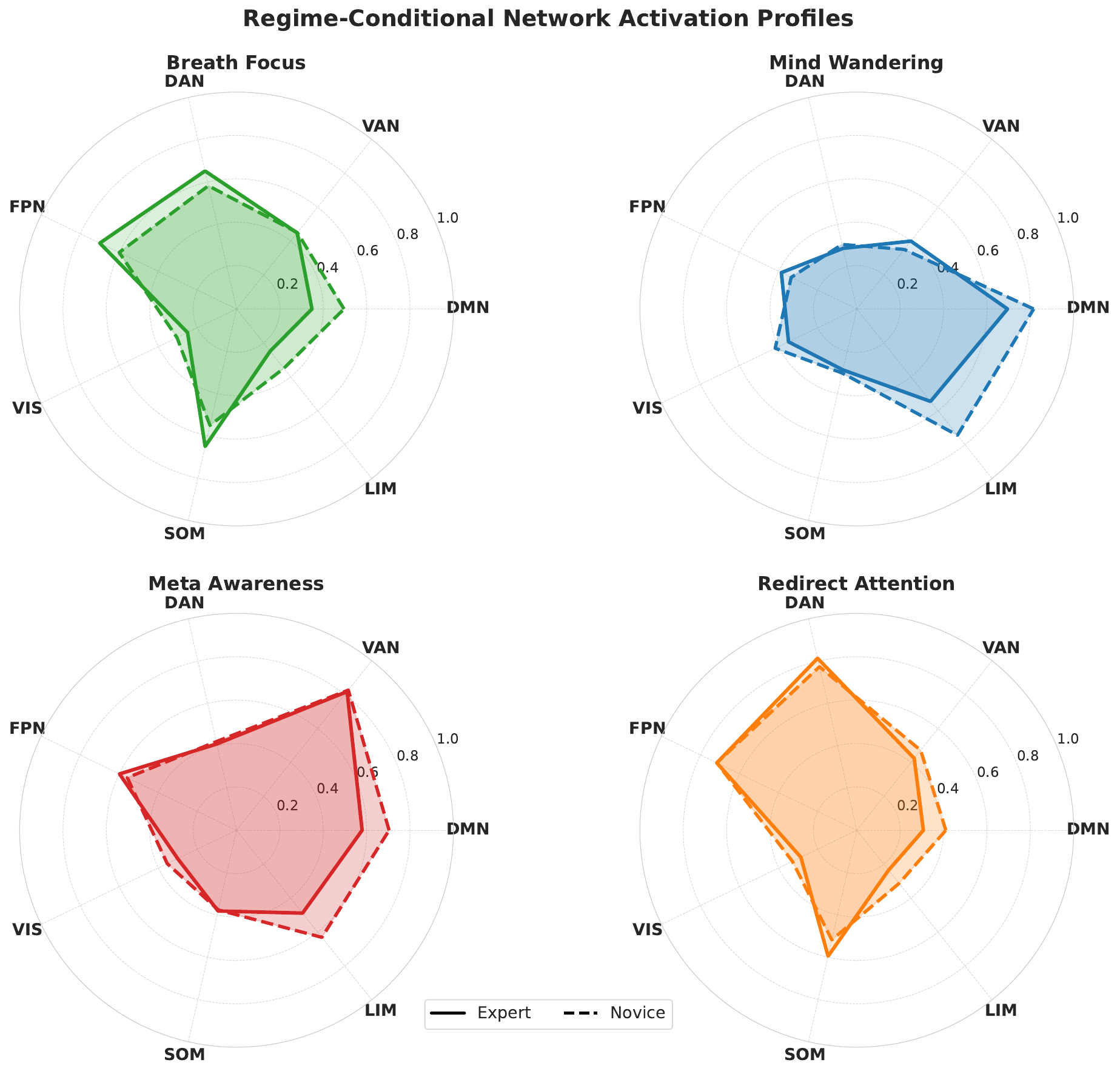} \\
\small{(B)} & \includegraphics[width=0.65\textwidth]{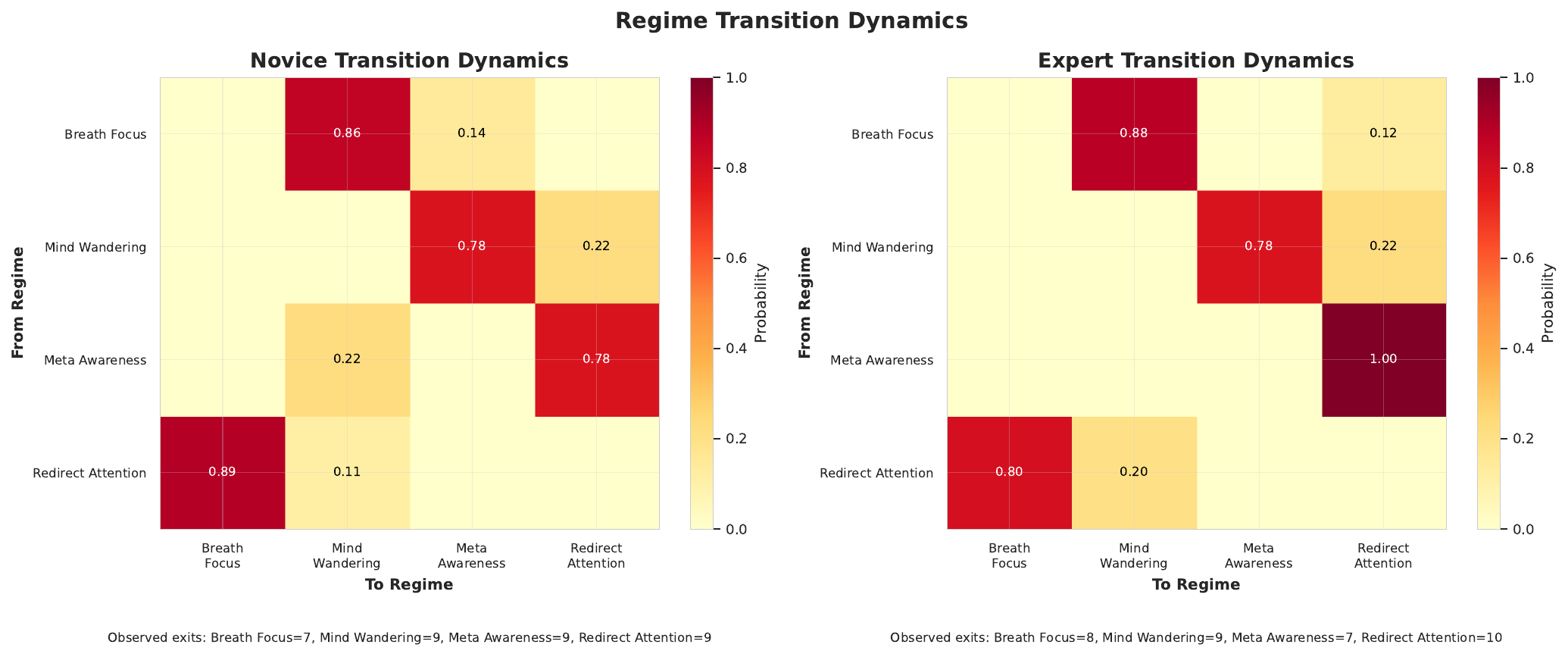} \\
\end{tabular}
\caption{\textbf{Regime-conditional profiles and transitions.} Values are descriptive summaries over the evaluation window of one seed-\StatsSeed{} run per phenotype. \textbf{(A) Network activation profiles.} Regime-conditional average activations. \textbf{(B) Transition dynamics.} Conditional transition proportions, excluding self-transitions. Observed exit counts by source regime are printed below each panel.}
\label{fig_4}
\par\smallskip\noindent{\small Alt text: Panel A shows four radar charts of seven network activations, one per regime, with the expert as a solid line and the novice dashed. The expert has lower default mode and higher attention-network activation in Breath Focus. Panel B shows two heat maps of observed transition proportions between regimes for novice and expert, with exit counts printed beneath.}
\end{figure*}

Fig.~\ref{fig_4} summarizes regime-conditional profiles. The configured regime attractors set targets, and the plotted profiles are time averages of the realized stochastic trajectory. Over the evaluation window, the most probable regime under L2's belief $\rho_t$ matched the regime drawn by L1 on \ExpertRegimeBeliefPercent\% (expert) and \NoviceRegimeBeliefPercent\% (novice) of steps.

\textbf{Network activation profiles.} In Breath Focus, the realized expert and novice means are \ExpertBreathFocusDMN{} and \NoviceBreathFocusDMN{} for DMN, \ExpertBreathFocusDAN{} and \NoviceBreathFocusDAN{} for DAN, and \ExpertBreathFocusFPN{} and \NoviceBreathFocusFPN{} for FPN. The configured phenotype contrast represents reduced self-referential processing and sustained attention directed toward the breath \citep{Brewer2011,Hasenkamp2012,MacLean2010}. During Mind Wandering, DMN means are \ExpertMindWanderingDMN{} (expert) and \NoviceMindWanderingDMN{} (novice). During MA, VAN means are \ExpertMetaAwarenessVAN{} and \NoviceMetaAwarenessVAN{}. VAN elevation reflects the configured MA attractor, not detection, which also occurs in other regimes (Section~\ref{sec:detect}).

\textbf{Regime durations and transitions.} Realized mean Breath Focus dwells were longer and Mind Wandering dwells shorter in the expert, as configured. Supplementary Table~S2 sets the realized mean dwells of all four regimes beside the configured means $T_s$. Fig.~\ref{fig_4}B gives the observed transition proportions, which are descriptive and rest on small exit counts, printed beneath each panel. In this run RA$\rightarrow$BF was lower in the expert, as in the five-run median, and the MW$\rightarrow$MA and RA$\rightarrow$BF differences both varied in direction across runs (Supplementary Table~S4).

\begin{figure*}[tbp]
\centering
\begin{tabular}{@{}c@{}}
\includegraphics[width=0.74\textwidth]{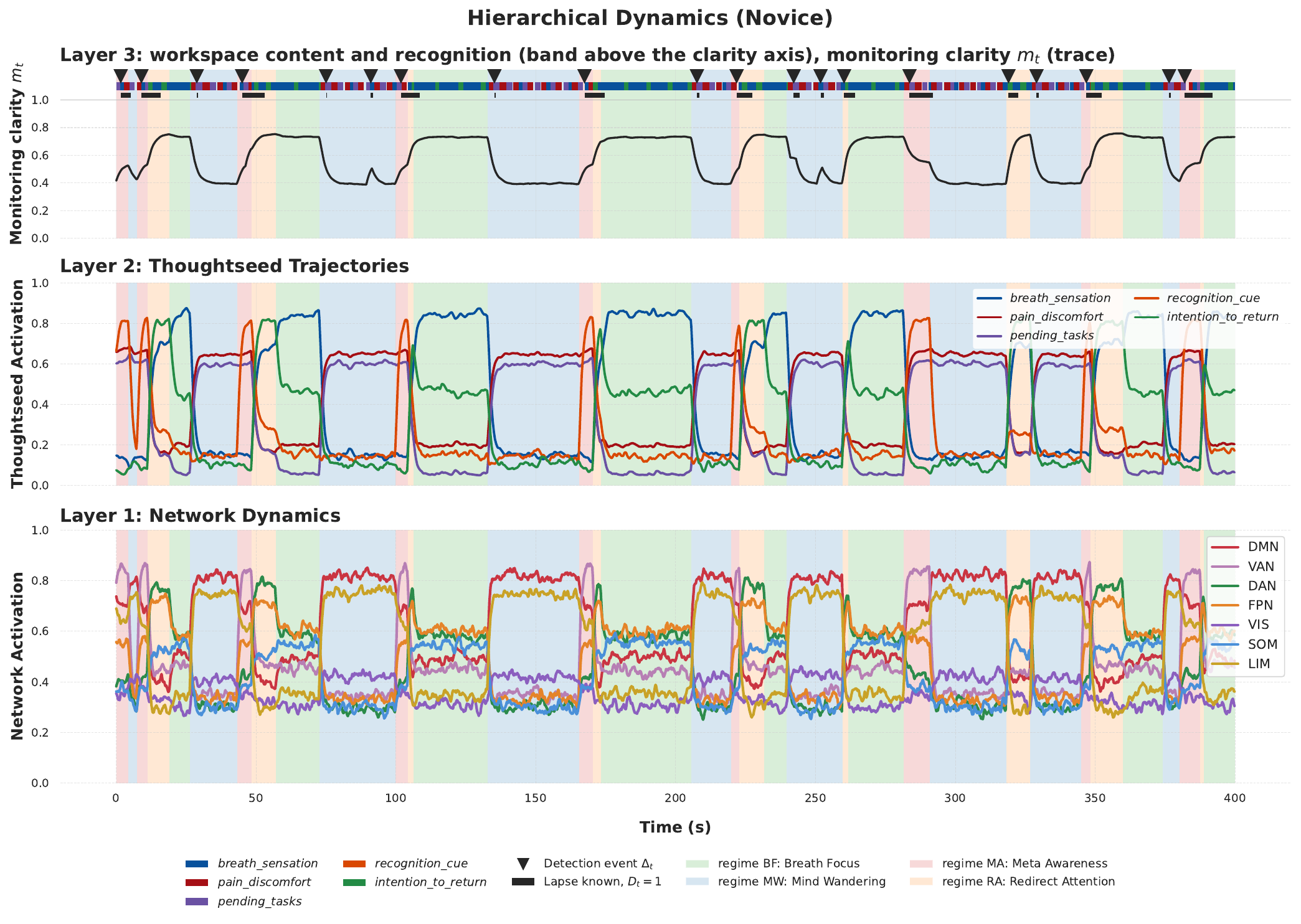} \\
\small{(A) Novice} \\[0.6em]
\includegraphics[width=0.74\textwidth]{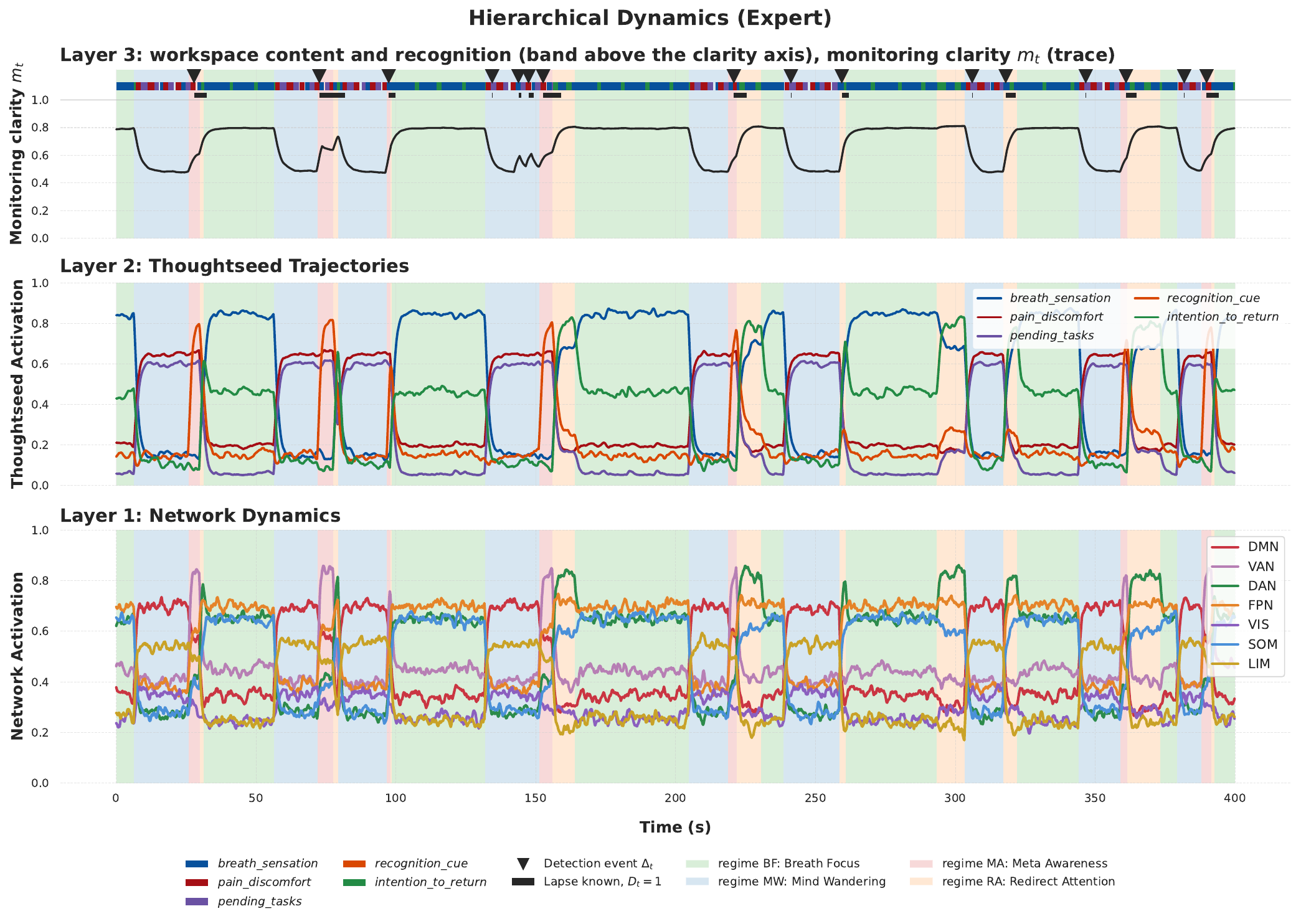} \\
\small{(B) Expert}
\end{tabular}
\caption{\textbf{Hierarchical dynamics with continuous thoughtseed trajectories.} One run per phenotype over the evaluation window, novice in (A) and expert in (B). The top row (L3) shows monitoring clarity $m_t$ with the accessed workspace content as a colored strip, with triangles at recognition onsets and bars where the lapse is known, the middle row (L2) shows the five continuous thoughtseed activations, and the bottom row (L1) shows the seven Yeo-network activations. The concurrent regime is shaded behind all three rows. Time is measured from the start of each run's evaluation window. All traces are on the activation scale $[0,1]$, and the L2 and L1 traces are smoothed for legibility by exponential moving averages ($\alpha=0.2$ and $0.3$).}
\label{fig_5}
\par\smallskip\noindent{\small Alt text: Two stacked panels, novice above and expert below, each with three rows. The rows show monitoring clarity with a colored strip of workspace content, recognition-onset triangles and recognition-interval bars, then five thoughtseed and seven network activation traces. Pale bands mark the regime. The expert spends more of the window in Breath Focus with higher monitoring clarity.}
\end{figure*}

\textbf{Workspace access, monitoring and recognition.} Fig.~\ref{fig_5} places the three layers on one timeline over the evaluation window. In the illustrated runs, regime changes in the network activations (L1) were accompanied by changes in the inferred thoughtseed activations (L2) and in workspace content (L3), and monitoring clarity followed with a delay. Clarity was typically high during Breath Focus, lower during Mind Wandering, and rose in Meta Awareness, where the \textit{recognition\_cue} activation was high. The breath sensation held the workspace on \ExpertBreathAccessPercent\% of steps in the expert run and \NoviceBreathAccessPercent\% in the novice run, with mean breath-access runs of \ExpertBreathRunSeconds{} and \NoviceBreathRunSeconds{}\,s. During Mind Wandering, distractor content held the workspace on \ExpertDistractorAccessMWPercent\% and \NoviceDistractorAccessMWPercent\% of steps. Mean monitoring clarity within Breath Focus was \ExpertClarityBF{} (expert) and \NoviceClarityBF{} (novice). These differences arise under shared access and monitoring rules, from the configured phenotype dynamics and what each phenotype learned under them. Recognition onsets and recognition intervals are marked above the clarity trace and are examined in Section~\ref{sec:detect}.

\subsection{Distraction, Recognition and Return}\label{sec:detect}
\textbf{Recognition and continued distraction.} Fig.~\ref{fig_5} shows recognition onsets outside MA, including brief recognitions early in Mind Wandering, while clarity carried over from Breath Focus was still above its resting level, and recognitions later in Mind Wandering. Across the five evaluation windows, most recognition onsets in Mind Wandering ended before any breath re-entry, as monitoring clarity declined. In Meta Awareness, most onsets were followed by breath re-entry before recognition offset, and a smaller number occurred with the breath already present. Supplementary Table~S5 gives the onset outcomes by regime. Recognition and the same distraction overlapped briefly in both phenotypes, with medians under a second (Table~\ref{tab:event_measures}). The proportion of distraction time with recognition was lower in the expert in the median paired difference, which varied in direction across runs (Supplementary Table~S4). Pooled over the five runs, a distractor held the workspace at \FiveRunExpertMetaAwarenessDetectionsDistractor{} of \FiveRunExpertMetaAwarenessDetections{} expert and \FiveRunNoviceMetaAwarenessDetectionsDistractor{} of \FiveRunNoviceMetaAwarenessDetections{} novice recognition onsets in MA.

\textbf{Recognition during established wandering.} With monitoring checks, recognition also began well after MW entry. Pooled over the five evaluation windows, recognition onsets during established Mind Wandering occurred at \FiveRunExpertEstablishedMWOnsetsPerMinute{} (expert) and \FiveRunNoviceEstablishedMWOnsetsPerMinute{} (novice) per minute of such time. Recognition remained intermittent. The share of complete episodes with no recognition onset after the first 3\,s was \FiveRunExpertEstablishedMWNoOnsetShare{} and \FiveRunNoviceEstablishedMWNoOnsetShare{}, and the 90th percentile of the longest stretch without recognition was \FiveRunExpertEstablishedMWLongestNinety{} and \FiveRunNoviceEstablishedMWLongestNinety{}\,s (counts in Supplementary Section~S4). The lower novice share at comparable rates is consistent with the longer configured novice Mind Wandering dwell, which gives episodes more opportunities for a check at the same per-step probability. It reflects the configured dwell difference and is not evidence about novice practitioners. A distractor held the workspace at \FiveRunExpertEstablishedMWDistractorShare{} (expert) and \FiveRunNoviceEstablishedMWDistractorShare{} (novice) of these onsets. A check requires a distractor at the check step, but recognition can begin on a later step, after the breath sensation has entered the workspace.

\begin{figure*}[tbp]
\centering
\includegraphics[width=0.95\textwidth]{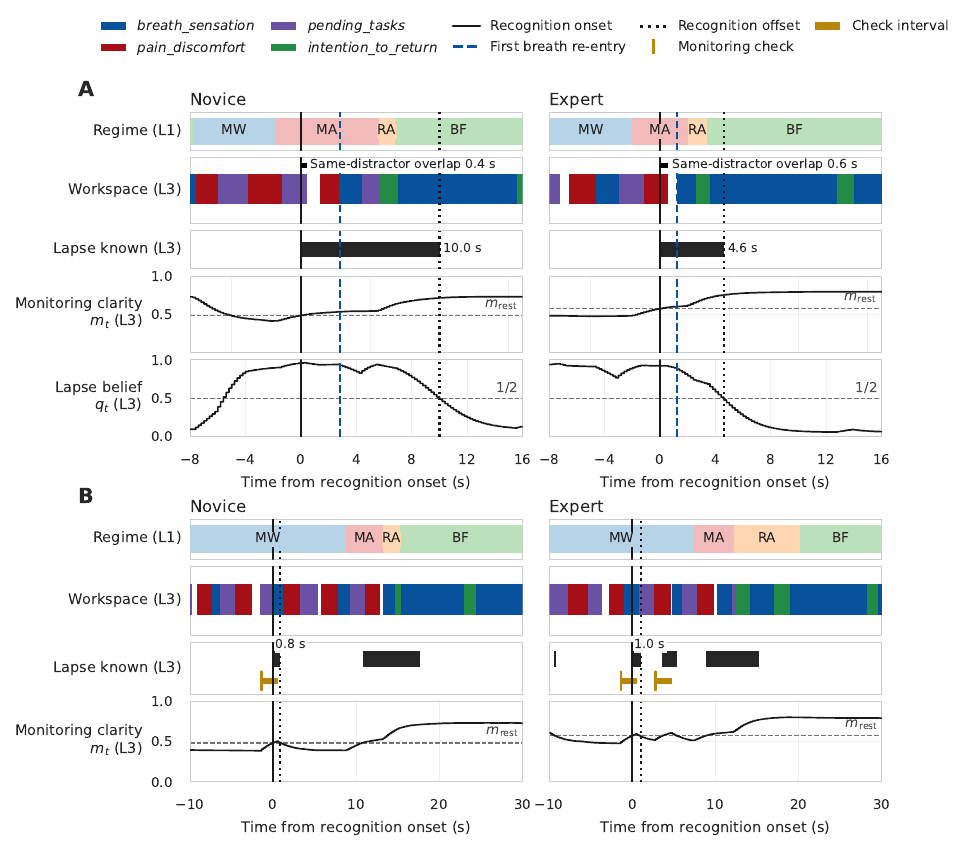}
\caption{\textbf{Selected recognition events.} Each panel shows one selected event per phenotype, novice left and expert right, aligned on recognition onset (0\,s), from the illustrative run (seed \StatsAnatomySeed{}, chosen by inspection of the five runs). The events are selected illustrations, not representative episodes. \textbf{(A)} The first recognition onset in MA in the window at which the same distractor holds the workspace from the step before onset to the step after it, selected for a lapse that runs from Mind Wandering through Meta Awareness and Redirect Attention to Breath Focus. Vertical lines mark recognition onset, the first breath re-entry and recognition offset, and the short black marker shows the same-distractor overlap. \textbf{(B)} The first recognition onset at least 3\,s after MW entry within a check interval, with nothing after the onset selected. Gold ticks mark monitoring checks, and gold bars mark the ten steps from each check, one monitor time constant, during which the check adds drive to monitoring clarity. Black bars mark intervals in which the lapse is recognized. A check interval beside an onset shows their timing, not that the check caused the onset. Panel B omits the lapse-belief track. The regime is L1's state, and the other tracks are computed in L3 without access to the regime.}
\label{fig_6}
\par\smallskip\noindent{\small Alt text: Two panels of novice and expert columns aligned on a recognition onset at time zero, with tracks for regime, workspace content, lapse known and monitoring clarity, and in panel A also lapse belief. In panel A, recognition begins in Meta Awareness while a distractor holds the workspace, the breath sensation re-enters, and Redirect Attention and Breath Focus follow. In panel B, a check occurs during Mind Wandering, the breath sensation enters the workspace, a brief recognition follows, and Mind Wandering continues before Meta Awareness, Redirect Attention and Breath Focus.}
\end{figure*}

\textbf{Selected trajectories.} Fig.~\ref{fig_6} shows two selected events per phenotype. Panel A shows recognition in Meta Awareness while the same distractor remains in the workspace, in a lapse selected because it ran from Mind Wandering through Meta Awareness and Redirect Attention to Breath Focus. In both examples recognition began while the distractor still held the workspace, so recognition did not displace it. After the distractor left, the breath sensation regained the workspace while the regime was still Meta Awareness, and Redirect Attention and Breath Focus followed. Recognition continued beyond the breath re-entry and ended early in Breath Focus, when the lapse belief fell below $\tfrac12$. It lasted longer in the novice example, in which a distractor re-entered once after the breath first returned. Panel B shows the first recognition onset during established Mind Wandering within a check interval, with nothing after the onset selected. In both examples a check occurred while a distractor held the workspace, the breath sensation then entered the workspace, and a brief recognition began while the regime was still Mind Wandering. Recognition ended as clarity fell after the check interval, and Mind Wandering continued until the configured exit to Meta Awareness, where a new recognition began. In the expert example, a second check was followed by another brief recognition. Breath re-entry and recognition offset provide separate endpoints for the measures summarized across runs in Table~\ref{tab:event_measures}.

\textbf{Return and recognition offset.} The median observed breath re-entry proportion was higher in the expert (Table~\ref{tab:event_measures}). Among eligible re-entries, the across-run median of within-run median recognition-to-breath latencies was lower in the expert, but the paired difference varied in direction. Regime-level return is a different endpoint. Over qualifying MA episodes with a direct exit to Redirect Attention, the across-run median recognition-to-RA latency was also lower in the expert, and the paired difference was negative in all five runs (Supplementary Table~S4). Among events with breath re-entry before or at recognition offset, recognition usually continued beyond that re-entry. Supplementary Section~S4 gives the time from the first breath re-entry to recognition offset and how often a distractor recurred in that interval.

\input{figures/event_measures.tex}

\subsection{Learning and Practice}\label{sec:practice}
\textbf{Breath access.} Training raised the breath-access prior from a uniform initial prior to \ExpertBreathAccessPrior{} and \NoviceBreathAccessPrior{}, and the learned access counts set resting clarity at \ExpertRestingClarity{} and \NoviceRestingClarity{}. Over training, the mean breath-access run lengthened in \FiveRunExpertBreathRunUp{} expert and \FiveRunNoviceBreathRunUp{} novice robustness runs, and the five-run median lengthened only in the expert (Supplementary Fig.~S2).

\textbf{Return policy.} In the illustrative runs, by the end of training the habit's probability of returning had reached \ExpertLearnedHabitReturn{} for the expert and \NoviceLearnedHabitReturn{} for the novice, from configured priors of 0.95 and 0.70. Over all MA exits in the window, the mean return effort in these runs was \ExpertReturnEffort{}\,nats for the expert and \NoviceReturnEffort{}\,nats for the novice. The expert-minus-novice return-effort difference was negative in all five paired runs (Supplementary Table~S4). The expert's habit assigns a higher probability to returning, so a return-favoring policy can remain closer to it. Between the first and last training blocks of \StatsPlotSteps{} steps, the mean return effort at MA exits fell from \ExpertReturnEffortFitFirst{} to \ExpertReturnEffortFitLast{}\,nats for the expert and from \NoviceReturnEffortFitFirst{} to \NoviceReturnEffortFitLast{}\,nats for the novice, and it fell in \FiveRunExpertReturnEffortDown{} expert and \FiveRunNoviceReturnEffortDown{} novice robustness runs (Supplementary Fig.~S2).

\FloatBarrier

\section{Discussion}

The central result is a proposed mechanism that dissociates experiencing a distraction from recognizing it. Detection is a separate L3 event that leaves the accessed content in place. In Meta Awareness, recognition can begin while a distractor still holds the workspace, and the distractor can then leave while the lapse remains recognized (Fig.~\ref{fig_6}). Same-distractor overlap ended either when clarity fell to its resting level or when the distractor left the workspace. Sustained recognition during ongoing, uninterrupted distraction remains transient in these runs, typically resolving in under a second (Table~\ref{tab:event_measures}). Recognition usually continued beyond breath re-entry (Section~\ref{sec:detect}). In the model, the breath becoming experienced again and sustained breath focus being judged re-established are distinct events. Their temporal separation has not been tested against experience.

Monitoring checks extend recognition into established Mind Wandering. Because a check is unscheduled and recognition follows only when the detection conditions hold, recognition came and went within established wandering while long stretches remained unrecognized (Section~\ref{sec:detect}), consistent with accounts in which meta-awareness of mind wandering is intermittent and wandering often continues unnoticed \citep{Schooler2002,Schooler2011}. Checks during established wandering are therefore unscheduled occasions for recognition, although a random check is not claimed to share the phenomenology of spontaneous thought. Recognition during wandering was brief and did not by itself bring return.

Recognition changes regime persistence and the choice made on an MA exit. An accessed distractor stabilizes Mind Wandering, while the learned habit supplies the default return choice \citep{WoodRunger2016}. These offer limited analogies with automatic constraints on thought and action in the dynamic framework of spontaneous thought \citep{Christoff2016}. Recognition releases stabilization and lets the MA policy depart from habit, a limited analogue of deliberate constraint. That departure was small because the habit already favored returning. Recognition does not force return, and return effort excludes its effect on exit timing. The fall of return effort over training (Section~\ref{sec:practice}) does not establish effortless focus \citep{Lutz2008}. This is an interpretive correspondence. The model does not measure spontaneity, and Mind Wandering onset is configured.

The results differ in how directly they follow from the configuration. The four-regime cycle and the phenotype dwell times, transition probabilities and network attractors are configured. L2 infers the regime with network signatures configured from the L1 attractors, and its belief matched the generated regime on nearly every step (Section~\ref{sec:verify}). Thoughtseeds therefore largely follow their regime-conditioned priors, and regime-level quantities, such as how often the breath holds the workspace, are strongly shaped by the configured regime structure. The timing of access and monitoring events is not fixed in advance but arises from the configured rules and each run's stochastic history. Monitoring checks occur with a configured probability, but whether and when recognition follows a check depends on clarity, the lapse belief and the workspace content. Where recognition begins, whether the breath re-enters before Redirect Attention, and how long same-distractor overlap lasts are outcomes. Detection itself requires that a distraction has been experienced in the current lapse, with clarity above its resting level and a lapse belief of at least one half, so the rule was designed to produce recognition-like events. Their occurrence is therefore not independent evidence that the model explains conscious recognition.

The expert-novice differences are consequences of the assumed phenotype settings. The two phenotypes share their update rules and every constant outside the phenotype tables. L2's starting decoder and transition model, L3's lapse-belief transitions and the habit prior are derived from those tables, and learned values diverge with training. Longer breath-access runs, higher monitoring clarity and lower return effort in the expert held in every paired run (Supplementary Table~S4). They are outcomes of the model under these assumptions, not independently established characteristics of meditation expertise. Other measures, such as the recognition-to-breath latency and some transition proportions, varied in direction across runs.

The present study advances the computational phenomenology of mental action in focused-attention meditation \citep{Ramstead2022,SandvedSmith2025} by relating first-person descriptions of meditative states to simulated network dynamics. The three-layer hierarchy (networks, thoughtseeds, metacognition) formalizes nested control over attentional contents, consistent with predictive-processing accounts \citep{LaukkonenSlagter2021}, classical descriptions of meditation \citep{Analayo2018} and accounts of self-organizing modes of brain function \citep{Safron2020}. By grounding L1 in the organization of large-scale brain networks \citep{Yeo2011} and motivating the hierarchy by the Neuronal Packet Hypothesis \citep{Ramstead2021,Yufik2019}, the model offers a biologically motivated implementation of competitive interactions among latent states. The hierarchy also orders timescales \citep{KiebelDaunizeauFriston2008}, though not strictly by layer (Section~\ref{sec:limitations}). That monitoring integrates more slowly than access selects is a modeling choice, motivated by dissociations between experience and awareness of that experience \citep{Schooler2002,DehaeneLauKouider2017}.

The hierarchy also separates automatic inference from monitoring-dependent regulation. L2 infers all candidate contents in the Helmholtzian sense of unconscious inference, which contributes to belief updating without the inferred content being accessed \citep{Hohwy2013,ParrPezzuloFriston2022}. Only the content selected into the L3 workspace is accessed, and monitoring bears on that content. This boundary offers a functional analogy with dual-process accounts, which contrast autonomous processes that generate default responses (System 1) with processing that draws on working memory to revise them (System 2) \citep{EvansStanovich2013,Kahneman2011}. L2's content inference is System-1-like, and regulation that depends on monitoring is System-2-like in its control function.

The deep parametric model of \citet{SandvedSmith2021}, from which the lapse belief is adapted, uses a meta-awareness level to set the precision of evidence about the attentional state. Here a monitoring check transiently adds drive to monitoring clarity, which weights evidence about the attentional state, but checks occur with a configured probability and are not selected for their epistemic value. The present model addresses a different question, how identifiable contents compete for workspace access and how a distinct recognition event relates to the experienced content. In recurrent-network models, shifts between focus and mind wandering arise within the networks' own dynamics \citep{Oyama2025,Idei2026}. Here the four-regime organization is specified instead.

\subsection{Empirical Correspondences and Measurement Relations}\label{sec:empirical}
Table~\ref{tab:empirical} sets each simulated signature against an empirical parallel and marks it as configured or emergent. A mechanism can be enabled by construction while its frequency, timing, and phenotype contrast remain outcomes to test.

\begingroup
\setlength{\tabcolsep}{4pt}
\setlength{\LTcapwidth}{\textwidth}
\begin{longtable}{>{\footnotesize\raggedright\arraybackslash}p{3.7cm}>{\footnotesize\raggedright\arraybackslash}p{4.1cm}>{\footnotesize\raggedright\arraybackslash}p{5.3cm}>{\footnotesize\raggedright\arraybackslash}p{1.6cm}}
\caption{\textbf{Empirical correspondences.} For each simulated signature, the mechanism that generates it, the empirical finding it parallels, and its epistemic status. \emph{Configured} signatures were built into the phenotype parameters or the shared rules and do not independently validate those assumptions. \emph{Emergent} signatures are outcomes not directly specified.}
\label{tab:empirical}\\
\hline
Simulated signature & Generating mechanism & Empirical parallel & Epistemic status \\
\hline
\endfirsthead
\hline
Simulated signature & Generating mechanism & Empirical parallel & Epistemic status \\
\hline
\endhead
Longer Breath Focus and shorter Mind Wandering dwells in experts (Supplementary Tables~S2 and~S4) & Phenotype dwell priors & Sustained-attention gains with training \citep{MacLean2010}. Reduced mind wandering with practice \citep{BrandmeyerDelorme2021,Tang2015} & Configured \\
Lower expert DMN, higher DAN/FPN during focus (Fig.~\ref{fig_4}A) & Phenotype network attractors $\boldsymbol{\mu}_x(s)$ & Reduced DMN engagement in experienced meditators \citep{Brewer2011}. Greater attention-related activation in long-term practitioners than in novices, with an inverted-U across practice hours \citep{BrefczynskiLewis2007} & Configured \\
VAN elevation in Meta Awareness (Fig.~\ref{fig_4}A) & MA attractor profile & Salience-network activity at awareness of mind wandering, approximated here by the VAN variable \citep{Hasenkamp2012,Seeley2007} & Configured \\
Transitions favor the MW$\rightarrow$MA$\rightarrow$RA$\rightarrow$BF route (realized proportions in Fig.~\ref{fig_4}B) & Configured dwell-time distributions and transition table. Stabilization changes exit timing, and in MA the return policy changes exit timing and successor choice & Four-phase cycle structure \citep{Hasenkamp2012}. Attentional regulation \citep{BrandmeyerDelorme2021} & Configured \\
Recognition onset while a distractor holds the workspace, in MA and in other regimes (Fig.~\ref{fig_6}, Section~\ref{sec:detect}), with brief same-distractor overlap (Table~\ref{tab:event_measures}) & Detection rule of Eq.~\eqref{eq:detect}, which leaves the accessed content in place & Self-caught reports of mind wandering \citep{Schooler2011}. Reported awareness of mind wandering during breath focus \citep{Hasenkamp2012} & Overlap permitted by construction. Observed frequency and duration emergent \\
Longer breath-access runs in the expert (Fig.~\ref{fig_5}, Supplementary Table~S4 and Fig.~S2) & Configured phenotype dwells, capacity-limited access with reverberation and adaptation, and shared access learning & GNW access and sustained broadcast \citep{Mashour2020,DehaeneChangeux2011}. Improved sustained attention with training \citep{MacLean2010} & Emergent \\
Higher monitoring clarity during Breath Focus in the expert (Fig.~\ref{fig_5}, Supplementary Table~S4) & Slow monitor driven by inferred first-order activations and the recognition cue, with resting level learned from accessed-content counts & Experts report more accurate introspection \citep{Fox2012} and greater meta-awareness \citep{Levinson2014}. How these measures map onto $m_t$ remains open & Emergent \\
Intermittent recognition during established Mind Wandering, with long unrecognized stretches (Section~\ref{sec:detect}) & Checks are eligible while a distractor holds the workspace. A check transiently adds drive to monitoring clarity, and detection follows only when its conditions hold & Intermittent meta-awareness during mind wandering, which often continues unnoticed \citep{Schooler2002,Schooler2011} & Check probability configured. Recognition timing and frequency emergent \\
Distractor access without detection & Access has no monitoring threshold, and distractors lower the monitor's drive & Experience without meta-consciousness \citep{Schooler2002}. Mind wandering and working memory \citep{Levinson2012} & Configured \\
\hline
\end{longtable}
\endgroup

Three distinctions shape comparison with experience. First, recognition onset need not coincide with departure from Breath Focus or entry into MA \citep{Schooler2011,Hasenkamp2012}. Second, return to the breath and entry into the Redirect Attention regime describe different aspects of recovery and need not occur together (Table~\ref{tab:event_measures}), and faster regime-level return did not mean uniformly faster return to the breath (Section~\ref{sec:detect}). Third, the proportion of distraction time with recognition is not an expertise scale. Its median paired difference was negative but varied in direction across runs, and its numerator and denominator both differ between the phenotypes, for example through the expert's higher breath access, so its mechanism remains unresolved.

The expert's higher monitoring clarity arises here through phenotype-dependent inferred activations and learned access counts, with shared monitor parameters. Empirical work suggests that expertise also improves monitoring while on task \citep{Fox2012,Levinson2014,Baird2014}, although these measures index introspective accuracy and reported meta-awareness rather than monitoring clarity as defined here. The model is therefore a baseline in which expertise acts on monitoring only through these routes.

The quantities in Table~\ref{tab:event_measures} are model-generated comparisons, and their experimental counterparts are proposed measurement relations, not equivalents. Self-caught reports or button presses mark reports of noticing \citep{Hasenkamp2012}, but comparison with detection and later regime transitions requires procedures that identify recognition and redirection boundaries \citep{Hasenkamp2012,Lutz2025}. Thought probes can compare accessed content with reported breath-related or distracting content, and can ask whether distraction was accompanied by awareness of mind wandering \citep{Schooler2011}, which parallels the proportion of distraction time with recognition (Supplementary Table~S4). Resumed breath-related report or breath counting \citep{Levinson2014} could time the return to the breath after reported recognition, although breath counting does not measure the belief threshold. Same-distractor overlap and the recognition interval have no validated measurement counterpart (Supplementary Table~S6). At the network level, state-space models such as GLHMMs \citep{Vidaurre2025} fitted to meditation fMRI can estimate network-state occupancy and transitions, and neural markers of conscious access offer a candidate comparison for workspace events \citep{Mashour2020}. Neither uniform expert superiority nor a fixed order of recognition, intention and return is required for these comparisons to be informative. These mappings require an observation model linking simulation variables to measurements, and the present runs remain descriptive.

\subsection{Limitations and Future Directions}\label{sec:limitations}
The workspace is a deliberately reduced form of GNW, a discrete-time, thresholded winner-take-one rule with reverberation and adaptation in place of a network of spiking populations. It is also functional. Network analyses identify a workspace with a functional rich club of hub regions that integrates perceptual, memory, evaluative and attentional information \citep{Deco2021}, whereas the model treats access as a property of the workspace slot, independent of any particular L1 network. Report, episodic memory, and elaboration of accessed content lie outside the model.

Network evidence, weighted by the learned observation precision, changes the corrected activations entering access competition, but workspace content still tracks the L1-generated regime closely. How much of access reflects reverberation, adaptation and learning rather than the regime structure remains to be separated in controlled runs. The access-credit rule rewards only the breath sensation, and other learned access biases, such as habitual rumination, lie outside the model. Relatedly, the five thoughtseeds modeled here are a fixed context-active subset of a larger dormant inventory. The present implementation represents only the active subset, so the restriction to five is imposed by construction.

The four-phase organization follows \citet{Hasenkamp2012}, whereas the dwell-time distribution and transition table are configured here, and entry into MA follows the transition table. L3's lapse belief does not cause that entry, and letting it do so is left to future work. Entry into Mind Wandering follows the configured dwell and transition process. In the saved runs, distractors reached the workspace after that entry, so accessed distracting content did not initiate these lapses. The MA priors, equal to those of MW except for \textit{recognition\_cue}, and the cue's term in the monitoring drive are specified rather than learned.

Detection is a discrete event computed in L3, and an explicit experiential report of recognition is not modeled. Whether detection events correspond to recognition in experience is untested. This uncertainty also applies to the brief recognitions early in Mind Wandering and to those during established Mind Wandering that followed monitoring checks. The belief's target, whether sustained breath focus is established, is approximated by the Breath Focus regime, and its transition probabilities by summed dwell means, which are not shown to aggregate the regime process accurately. The modeled timing relationships among breath access, experienced intention and the belief that sustained focus is established remain unvalidated against experience. The release of stabilization during recognition is a modeling assumption that awaits empirical calibration.

Monitoring checks are configured opportunities for monitoring. Their probability is shared by both phenotypes, calibrated for illustration and not learned with practice. The ten-step check interval is anchored at one monitor time constant. Recognition during established Mind Wandering is an internal event and is not equivalent to a self-caught report. Recognition begun during Mind Wandering can last until the configured exit, but the return policy acts only in MA, so the model does not represent deliberate return from Mind Wandering.

The architecture is motivated by active inference and composed of separate rules for inference, access, monitoring and control. An active-inference-inspired one-step score affects MA successor choice and exit odds. Staying is scored by the regime it continues, recognition acts only through the precision, and the stabilization remains a rule. L2's network signatures are configured because, in exploratory fits from a random start, its decoder separated the regimes but did not identify which was which. Its transition model assumes geometric dwells and omits the Gamma dwell-time distribution and the descending control. L2 learns a single, state-independent observation precision. State-dependent precision, epistemic value, long-horizon planning and deeper hierarchical structure are left to future work. Descending control is limited to stabilization and the forwarded return policy and exit-odds factor, without a descending prediction of network activity. L1 is the generative process for L2 and L3 by construction, and the model does not establish that large-scale network activity is separated from content inference by a Markov blanket in the brain \citep{Hipolito2021}.

Parameter choices (e.g., dwell distributions), while theoretically motivated, should be calibrated against empirical measures (e.g., breath-count accuracy, experience sampling) within neurophenomenological paradigms \citep{Varela2017,Lutz2025}, as a recent review of active-inference models of meditation also urges \citep{Tal2026}. The layer separation is one configured factor of two (L1's slowest relaxation mode at most 0.5\,s, L2 1\,s, L3 2\,s), modest beside the roughly order-of-magnitude separation in the illustrative hierarchy of \citet{KiebelDaunizeauFriston2008}, so the layers overlap in time. The sensitivity of the reported dynamics to this factor remains untested.

A next phase could extend the model toward fully autonomous, interacting thoughtseeds, each with its own generative dynamics and precision control. This would enable emergent coordination via multi-agent interactions and dynamic Markov-blanket detection \citep{BeckRamstead2025}. A dose--response analysis, varying practice exposure with configured phenotype settings held fixed, would examine how training duration changes workspace dynamics. A further extension would let expertise act on the monitor itself while on task, for example through a model in which checking is selected for its expected information gain and practice changes that selection. Finally, L2's transition model could be learned and extended to represent elapsed regime duration and descending control, with the four-phase organization retained as a structural prior.

\section{Conclusions}

We have described a three-layer architecture that realizes the lapse-and-recognition dynamics of the four-phase canonical cycle of focused attention. Mental content, represented by thoughtseeds (latent causes with configured priors and a learned mapping to network activity), gains functional access through a capacity-limited workspace in which one content is held at a time. A slower graded monitor biases access, enters detection, weights the evidence about the attentional state and, while the lapse is recognized, sets the precision of the return policy in MA. The architecture draws selected commitments from active inference, global workspace theory, dual-process accounts of cognition and Buddhist analyses of mind, with explicit interfaces between adjacent layers.

The architecture's central distinction is between ongoing distraction and its recognition. Recognition is represented by an indicator in L3, separate from entry into Meta Awareness. In the simulations, recognition could begin while the same distractor remained in the workspace, and this overlap was usually brief. With monitoring checks, recognition also began intermittently during established Mind Wandering, while long stretches of wandering remained unrecognized. Recognition releases the stabilization that prolongs the current regime and lets the MA return policy depart from its learned habit, a limited analogy with automatic and deliberate constraints on thought. Retaining the four-regime organization preserves comparability with the level at which the cycle is reported empirically, and adding the graded monitor and detection preserves a distinction that the regimes alone would lose.

Throughout, we have separated what was configured, including entry into Mind Wandering, from what the coupled mechanisms produced (Table~\ref{tab:empirical}). Among the outcomes is the expert's higher monitoring clarity, which arises through phenotype-dependent inferred activations and learned access counts, with shared monitor parameters.

The workspace models functional access, leaving phenomenal consciousness outside its scope. Within that bound, the architecture provides a tractable account of how experienced content, the detection of distraction and sustained monitoring can play distinct and separable roles in regulating attention, and makes their relationships explicit for simulation and empirical comparison. Because access, monitoring, detection and return are shared mechanisms, while the regimes, contents, transition tables and network attractors are configured, the architecture could be reconfigured for practices with other objects, including those of other contemplative traditions, and for lapses of attention in everyday tasks. Such models would bring first-person descriptions from different practices and tasks into one computational framework, and so contribute to the study of conscious access and metacognition.

\section*{Conflict of Interest Statement}
The authors declare that the research was conducted in the absence of any commercial or financial relationships that could be construed as a potential conflict of interest.

\section*{Author Contributions}
Conceptualization and methodology: P.C.K.; software: P.C.K., D.A.F.; writing (original draft): P.C.K.; writing (review and editing): P.C.K., D.A.F., G.P.; supervision: G.P.; funding acquisition: G.P. All authors have read and agreed to the current version of the manuscript.

\section*{Funding}
This research was partially funded by grant PID2021-122136OB-C22 from \url{MICIU/AEI/10.13039/501100011033} and by ERDF ``A way of making Europe'', and by the AGAUR research support grant 2021 SGR 01035 from the Department of Research and Universities of the Generalitat de Catalunya (G.P.).

\section*{Acknowledgments}
We thank the organizers and participants of IWAI 2025 (6th International Workshop on Active Inference), where an earlier stage of this work was presented \citep{Kavi2026}. Their feedback helped develop it from an initial rules-based approach into the present model, improving both the conceptual architecture and the mathematical model.

\section*{Ethics Statement}
This study reports a computational simulation model and did not involve human participants, human data, or animal subjects. No ethical approval was required.

\section*{Supplementary Material}
The Supplementary Material contains the model specification (Section~S1), parameter tables (Section~S2), outcome-measure definitions (Section~S3), training diagnostics, practice curves and five-run summaries (Section~S4), and the contemplative foundations of the architecture (Section~S5).

\section*{Data Availability Statement}
All simulation code, configuration files, and reference parameters used in this study are openly available at \url{https://github.com/prakash-kavi/thoughtseeds_model}.

\end{document}


\singlespacing
\InputIfFileExists{figures/stats.tex}{}{C
  \InputIfFileExists{stats.tex}{}{%
    \errmessage{stats.tex not found: upload figures/stats.tex or place stats.tex alongside supplementary.tex}%
  }%
}
\InputIfFileExists{figures/robustness_macros.tex}{}{C
  \InputIfFileExists{robustness_macros.tex}{}{%
    \errmessage{robustness_macros.tex not found: upload figures/robustness_macros.tex or place it alongside supplementary.tex}%
  }%
}

\def\ExpectedModelVersion{clarity-v48}
\ifdefined\StatsModelVersion
  \edef\ActualModelVersion{\StatsModelVersion}
  \ifx\ActualModelVersion\ExpectedModelVersion\else
    \errmessage{Statistics use an older model: rerun simulations and scripts/export_stats.py}
  \fi
\else
  \errmessage{Statistics predate MA detection: rerun simulations and scripts/export_stats.py}
\fi
\section*{Supplementary Material}
\noindent\textit{Recognizing Mind Wandering: A Computational Phenomenology of Focused-Attention Meditation}

\medskip
\noindent This supplement provides the model specification, parameter tables, outcome-measure definitions, supplementary results and the contemplative foundations of the architecture.
\setcounter{secnumdepth}{2}
\setcounter{section}{0}
\setcounter{subsection}{0}
\setcounter{equation}{0}
\renewcommand\thesection{S\arabic{section}}
\renewcommand\thesubsection{S\arabic{section}.\arabic{subsection}}

\setcounter{figure}{0}
\renewcommand\thefigure{S\arabic{figure}}

\setcounter{table}{0}
\renewcommand\thetable{S\arabic{table}}

\makeatletter
\@addtoreset{equation}{subsection}
\makeatother
\renewcommand\theequation{S\arabic{section}.\arabic{subsection}.\arabic{equation}}

\section{Model Specification}

\subsection{Initialization and Numerical Conventions}

Symbols follow Table~1 of the main text, and main-text equations are cited by their numbers. The frozen rollout keeps the five learned quantities and resets to BF with zero dwell age, network activations at $0.5$, latent activations at the clipped BF prior, an empty workspace, resting monitoring clarity, and cleared adaptation, detection, departure evidence and pending credit. The belief $q_t$ starts at its no-evidence stationary value $\varepsilon_{\mathrm{on}}/(\varepsilon_{\mathrm{on}}+\varepsilon_{\mathrm{off}})$ (Eq.~\eqref{eq:supp:belief_predict}), and L2's regime belief $\rho_t$ starts in BF. $\mathbf{x}_t\in\mathbb{R}^{7}$ and $\mathbf{z}_t\in\mathbb{R}^{5}$ are clipped element-wise to $[0.05,0.9]$ ($\mathbf{x}_t$ after each Euler substep), and $m_t,r_t\in[0,1]$.

\subsection{Step 1, Network Dynamics and Regime (L1)}

\paragraph{Switching multivariate OU.}
L1 is the switching multivariate OU process of Eq.~1 in the main text, with $\mathbf{W}_t$ a 7D Wiener process and diffusion covariance $\sigma_x^2I$, integrated by Euler--Maruyama with two substeps per step. The attractor $\boldsymbol{\mu}_x(s)$ is the configured network profile of regime $s$ (\texttt{NETWORK\_PROFILES} in \texttt{config/profiles.py}).
\paragraph{Baseline duration model (L1).}
The baseline duration model uses the elapsed dwell time $a_t$ and a discrete Gamma dwell-time distribution \citep{Yu2010} with phenotype- and regime-specific mean $T_s$ and shared coefficient of variation $c_v=0.6$ (Table~\ref{tab:dwell_transitions}). Unlike the geometric dwells of a time-homogeneous hidden Markov model \citep{BastianSackur2013}, the baseline exit probability depends on elapsed dwell time. In steps, shape is $k=c_v^{-2}$ and scale is $(T_s/dt)/k$. Integer-age masses $f_s(a)$ are proportional to the Gamma density and normalized over a finite tail. The baseline exit probability at dwell age $a$ is $h_s(a)=f_s(a)/\sum_{j\ge a}f_s(j)$, the probability that the regime ends at its $a$-th step given that it has lasted $a-1$ steps. L1 evaluates it at $a=a_t+1$ on the update after step $t$.

\paragraph{Drift/coupling matrix.}
$\Theta(s)$ has a shared regime-specific off-diagonal template, with 34 nonzero entries given in \texttt{THETA\_BASE} in \texttt{config/defaults.py}. Each diagonal entry is the sum of the absolute off-diagonal entries in its row plus the settling margin $\theta_{\mathrm{set}}=\max[3/\min T_s,\,k_{\mathrm{sep}}/\tau_{\mathrm{dyn}}]$, with the minimum over BF, MW and MA in both phenotypes (Table~\ref{tab:core_params}). By the Gershgorin circle theorem \citep{HornJohnson2012}, every eigenvalue then has real part at least $\theta_{\mathrm{set}}$, so the process is mean-reverting even for asymmetric coupling \citep{Gilson2016}.

\paragraph{Timescale ordering.} Across all three layers, one-step workspace selection is fastest. The relaxation of L1, L2 and L3 follows, with L1's slowest mode at most 0.5\,s, $\tau_{\mathrm{dyn}}=1$\,s and $\tau_m=2$\,s, then content adaptation and the baseline regime dwells, in no fixed order between them, and learning of the access prior is slowest. The settling margin above enforces the separation of layer relaxation by the configured factor $k_{\mathrm{sep}}$.

\subsection{Step 2, Content Inference (L2)}

\paragraph{Decoder start.}
The decoder $W\mathbf{z}+\mathbf{b}$ starts from configured network signatures, the minimum-norm affine map $(W_0,\mathbf{b}_0)$ with $W_0\boldsymbol{\mu}_z(s)+\mathbf{b}_0=\boldsymbol{\mu}_x(s)$ for every regime, obtained with the pseudoinverse of the mean-centred regime priors (Table~\ref{tab:thoughtseed_priors}) and network attractors of the phenotype. This start supplies the regime labels, and the decoder is then learned (\S S1.6). With this initialization, L2's most probable regime matched the regime generated by L1 on nearly every step of the evaluation window (Section~3.1 of the main text). This demonstrates regime recovery under configured network signatures. It does not establish that the intended regime labels can be learned without that initialization.

\paragraph{Prediction and correction.}
For each regime $s$, L2 predicts the five thoughtseed activations by one Euler step of an OU process toward $\boldsymbol{\mu}_z(s)$ with time constant $\tau_{\mathrm{dyn}}$ and the L1 diffusion variance rate $\sigma_x^2$,
\begin{equation}
\begin{aligned}
\mathbf{z}_{t,s}^{-} &= \mathbf{z}_{t-1}^{*} + \frac{dt}{\tau_{\mathrm{dyn}}}\bigl(\boldsymbol{\mu}_z(s)-\mathbf{z}_{t-1}^{*}\bigr), \\
P_t^{-} &= \phi^2 P_{t-1} + Q\,I,\qquad \phi=1-dt/\tau_{\mathrm{dyn}},\quad Q=\sigma_x^2\,dt, \\
\rho_t^{-}(s) &= \textstyle\sum_{s'} B(s\mid s')\,\rho_{t-1}(s').
\end{aligned}
\label{eq:supp:z_predict}
\end{equation}
The transition model has $B(s'\mid s')=1-dt/T_{s'}$, with $T_{s'}$ the configured dwell mean (Table~\ref{tab:dwell_transitions}), and splits the exit probability $dt/T_{s'}$ over successors by the transition table (Table~\ref{tab:dwell_transitions}). This geometric-dwell approximation omits L1's Gamma dwell-time distribution and descending control. The belief is updated by how well each regime's prediction explains the network activity,
\begin{equation}
\rho_t(s) \propto \rho_t^{-}(s)\,\mathcal{N}\bigl(\mathbf{x}_t;\,W\mathbf{z}_{t,s}^{-}+\mathbf{b},\,WP_t^{-}W^{\top}+\Pi_x^{-1}I\bigr).
\label{eq:supp:regime_belief}
\end{equation}
The negative log of its normalizer is L2's surprisal, shown in Supplementary Fig.~\ref{fig:S1}. Each regime's prediction is then corrected by minimizing over the activations $\mathbf{z}$ the objective
\begin{equation}
F_{t,s}(\mathbf{z}) = \underbrace{\Pi_x\lVert\mathbf{x}_t-(W\mathbf{z}+\mathbf{b})\rVert^2}_{\text{reconstruction error}} + \underbrace{(\mathbf{z}-\mathbf{z}_{t,s}^{-})^{\top}\Pi_{z,t}(\mathbf{z}-\mathbf{z}_{t,s}^{-})}_{\text{prior deviation}},
\label{eq:supp:l2}
\end{equation}
where $\Pi_{z,t}=(P_t^{-})^{-1}$ is the prior precision. Each unconstrained minimizer is a Kalman measurement update, with posterior covariance $P_t=(\Pi_xW^{\top}W+\Pi_{z,t})^{-1}$. L2 combines the corrected estimates by a first-order generalized pseudo-Bayesian (GPB1) collapse \citep{BarShalom2001}. The implemented estimate $\mathbf{z}_t^{*}$ is the belief-weighted mean of the corrected estimates, clipped componentwise to $[0.05,0.9]$. By moment matching, the carried covariance combines the within-regime covariance with the between-regime spread around the unclipped mean. Clipping changes the mean without adjusting that covariance. The carried covariance starts, and restarts in the frozen rollout, at the no-evidence fixed point $Q/(1-\phi^2)\,I$. Neither L2 precision has monitoring clarity as a direct input.

\subsection{Steps 3 to 5, Access, Monitoring, Detection and Descending Control}

\paragraph{Content quantities.}
The quantities below are computed from the configured thoughtseed priors $\mu_{z,i}(s)$, and their values are in the lower block of Table~\ref{tab:thoughtseed_priors}. The \emph{diagnosticity} of thoughtseed $i$ for regime $s$ is its prior normalized across regimes, $d_i(s)=\mu_{z,i}(s)/\sum_{s'}\mu_{z,i}(s')$. The \emph{on-task contrast} $c(i)$ compares a content's prior in MW, treated as off-task for this contrast, with its mean over the other three regimes, and the \emph{lapse evidence} $\lambda(o)$ of a first-order content compares its mean prior over MW, MA and RA with its prior in BF,
\begin{equation*}
c(i)=\ln\frac{\tfrac13\sum_{s\neq\mathrm{MW}}\mu_{z,i}(s)}{\mu_{z,i}(\mathrm{MW})},\qquad
\lambda(o)=\ln\frac{\tfrac13[\mu_{z,o}(\mathrm{MW})+\mu_{z,o}(\mathrm{MA})+\mu_{z,o}(\mathrm{RA})]}{\mu_{z,o}(\mathrm{BF})}.
\end{equation*}
The \emph{on-task weight} $w_i=\tanh[c(i)/2]\in[-1,1]$ grades the access bias. An empty slot has $\lambda(\varnothing)=0$, and \texttt{recognition\_cue} supplies no lapse evidence.

\paragraph{Step 3, access history (L3).}
The access-history term of Eq.~2 of the main text is $\xi_{i,t}=\kappa\,\delta_{i,o_{t-1}} - w_a\,r_{i,t-1}$, the reverberation bonus of the incumbent $o_{t-1}$, with $\delta$ the Kronecker delta, minus the adaptation of content $i$. Adaptation follows occupancy, $r_{i,t}=r_{i,t-1}+(dt/\tau_a)(\delta_{i,o_t}-r_{i,t-1})$. With $w_a=z_{\max}+\kappa$, where $z_{\max}$ is the upper activation bound, saturated adaptation cancels the largest activation and the reverberation bonus.

\paragraph{Step 4, monitoring contrast (L3).}
The contrast in Eq.~3 of the main text is
\begin{equation}
\bar c_t=\frac{\sum_{i\in\mathcal{A}} z^*_{i,t}\,c(i)}{\sum_{i\in\mathcal{A}} z^*_{i,t}}+c(\mathrm{rc})\bigl(z^*_{\mathrm{rc},t}-\mu_{z,\mathrm{rc}}(\mathrm{MW})\bigr),
\label{eq:supp:contrast}
\end{equation}
where $\mathcal{A}$ is the set of the four first-order thoughtseeds and rc denotes \texttt{recognition\_cue}. The first term is the activation-weighted mean on-task contrast, and the second is the cue's activation above its MW prior, weighted by its own on-task contrast $c(\mathrm{rc})$.

\paragraph{Step 4, monitoring checks (L3).}
After access, a monitoring check occurs with probability $p_k\,\mathbb{1}[\lambda(o_t)>0]$, drawn from a random-number stream of its own, seeded with the run. The check-interval indicator of Eq.~3 of the main text is $e_t=\mathbb{1}[t-t_k<\tau_m/dt]$, where $t_k$ is the most recent check step, with $e_t=0$ before the first check. Each interval therefore covers ten steps including the check step, and a new check restarts it. A check reaches access only from the next step, through $m_t$. The check contrast is the second term of Eq.~\eqref{eq:supp:contrast} at the cue's MA prior, $c_k=c(\mathrm{rc})\bigl(\mu_{z,\mathrm{rc}}(\mathrm{MA})-\mu_{z,\mathrm{rc}}(\mathrm{MW})\bigr)\approx0.715$. The interval length of one monitor time constant is a design anchor, not a derivation. With $dt/\tau_m=0.1$, ten steps close about two thirds of the gap between $m_t$ and a fixed target. The probability $p_k$ is configured. It was calibrated to the check count of an alternative illustrative mechanism, in which checks followed distractor offsets, and it is not fitted to data. While a distractor remains accessed, it gives on average one check per about 19\,s of accumulated distractor access. This is not the mean time between checks across a run, which also depends on how often distractors hold the workspace.

\paragraph{Step 4, lapse belief and detection (L3).}
The predicted belief of Eq.~4 of the main text is
\begin{equation}
q_t^{-}=(1-\varepsilon_{\mathrm{off}})\,q_{t-1}+\varepsilon_{\mathrm{on}}(1-q_{t-1}),\qquad
\varepsilon_{\mathrm{on}}=dt/T_{\mathrm{BF}},\quad
\varepsilon_{\mathrm{off}}=dt/(T_{\mathrm{MW}}+T_{\mathrm{MA}}+T_{\mathrm{RA}}),
\label{eq:supp:belief_predict}
\end{equation}
where $\varepsilon_{\mathrm{on}}$ and $\varepsilon_{\mathrm{off}}$ are dwell-based approximations of the per-step probabilities that sustained breath focus ends and that it is re-established. Grouping RA with MW and MA is an operational approximation, since breath focus can begin during RA and distractors can hold the workspace during BF. The dwell-based $\varepsilon_{\mathrm{off}}$ assumes one passage through MW, MA and RA, and $\lambda(o)$ weights the three regimes equally, so both are approximations. For two states, Eq.~4 of the main text is the attentional-state update of \citet[Eq.~1]{SandvedSmith2021} without its ascending term, written in log-odds, with $m_{t-1}\,dt/\tau_{\mathrm{dyn}}$ in the role of the likelihood precision, which a discrete higher-level state sets in that model.

The departure-evidence flag of Eq.~5 of the main text is updated as $R_t=\max\bigl(\mathbb{1}[\lambda(o_t)>0],\,R_{t-1}\,\mathbb{1}[q_t\ge\tfrac12]\bigr)$, with $D_0=R_0=0$ after reset. Content accessed at step $t$ can support an event at step $t$.

\paragraph{Step 4, return-policy score (L3).}
The score $G(\pi)$ of Eq.~6 of the main text, for the regime $s_\pi$ entered by $\pi$, is the outcome-risk-plus-ambiguity form of expected free energy \citep{DaCosta2020},
\begin{equation}
G(\pi)=D_{\mathrm{KL}}\bigl[\widehat P(\cdot\mid s_\pi)\,\|\,C\bigr]+H\bigl[\widehat P(\cdot\mid s_\pi)\bigr]=-\textstyle\sum_o \widehat P(o\mid s_\pi)\ln C(o),
\label{eq:supp:policy}
\end{equation}
which reduces to cross-entropy under the one-step approximation that each candidate selects a definite successor regime with a fixed content forecast, so that ambiguity equals forecast entropy. The forecast $\widehat P(o\mid s_\pi)=\mu_{z,o}(s_\pi)/\sum_{o'}\mu_{z,o'}(s_\pi)$ and the preference $C(o)=d_o(\mathrm{BF})/\sum_i d_i(\mathrm{BF})$ are normalized over the four first-order contents, the outcomes that can hold the workspace. Here $\gamma_t$ is computed rather than inferred. The one-step policy provides no policy-specific observation before the exit choice, so there is no evidence that would update it.

\paragraph{Exit timing in MA.} $P(\pi)$ is the conditional, given an exit, of one distribution over staying in MA and the two exits. With the stabilized exit probability $\tilde h_t$ of Eq.~7 of the main text at $s_t=\mathrm{MA}$, its prior is $p_0(\mathrm{stay})=1-\tilde h_t$ and $p_0(\pi)=\tilde h_t\,E(\pi)$, and
\begin{equation}
\begin{gathered}
P(a)\propto p_0(a)\,e^{-\gamma_tG(a)},\qquad
\Pr(s_{t+1}\neq\mathrm{MA}\mid s_t=\mathrm{MA},a_t,S_t,\gamma_t)=1-P(\mathrm{stay})=\frac{\tilde h_t\Lambda_t}{1+\tilde h_t(\Lambda_t-1)},\\
\Lambda_t=\sum_\pi E(\pi)\,e^{-\gamma_t[G(\pi)-G(\mathrm{MA})]},
\end{gathered}
\label{eq:supp:policy_exit}
\end{equation}
where staying is scored for one step by $G(\mathrm{MA})$, which equals $G(\mathrm{MW})$ because MA and MW have the same priors for the first-order contents. Given an exit, the successor probabilities are $P(\pi)$ of Eq.~6 of the main text, and at $\gamma_t=0$, $\Lambda_t=1$ and the exit probability is $\tilde h_t$. With the configured scores and habits, higher precision increases the MA exit probability at fixed dwell age and stabilization when $0<\tilde h_t<1$, and the endpoints remain zero and one. Return effort (Section~2.3 of the main text) is the divergence of the successor policy $P(\pi)$ from the habit $E$, so it excludes the change in exit timing.

\subsection{Adjacent-Layer Interfaces}

Adjacent layers exchange only the signals below, an implementation motivated by Markov-blanket formulations \citep{Kirchhoff2018,Ramstead2021}. In this interface interpretation, ascending signals play the role of sensory states and descending signals that of active states. L1 sends $\mathbf{x}_t$ to L2, and L2 sends $\mathbf{z}_t^{*}$ to L3. L3 returns $o_t$, $D_t$, $P(\pi)$ and $\Lambda_t$, which L2 forwards to L1 in the same step as descending control, for use at L1's next update. L2's regime belief $\rho_t$ stays internal to L2, and $q_t$, $R_t$ and the detection event stay internal to L3.

\subsection{Learning}

Learning follows inference and control during training, subject to the update conditions below (Algorithm~1 of the main text). All five learned quantities are frozen in the rollout. Learning rates and the initial concentration $\alpha_0$ are in Table~\ref{tab:core_params}.

\paragraph{Learned decoder.}
With $\mathbf{z}_t^{*}$ held fixed, the decoder $(W,\mathbf{b})$ takes one stochastic gradient descent (SGD) step, with the gradient norm clipped, on the reconstruction error $L_t=\lVert\mathbf{x}_t-(W\mathbf{z}_t^{*}+\mathbf{b})\rVert^2$. The loss is not weighted by $\Pi_x$.

\paragraph{Learned observation precision.}
In the same step, the observation variance $v=\Pi_x^{-1}$ moves toward the squared error per network expected under the Gaussian posterior,
\begin{equation}
v \leftarrow v + \eta\,\Bigl(\bigl[L_t+\operatorname{tr}(WP_tW^{\top})\bigr]/D_x - v\Bigr),
\label{eq:supp:obs_precision}
\end{equation}
with learning rate $\eta$, $D_x=7$ networks, $L_t$ evaluated at $\mathbf{z}_t^{*}$, and $P_t$ the collapsed posterior covariance, which includes the spread of the regime-specific estimates. When clipping leaves the mean unchanged, the fixed point minimizes over $\Pi_x$ the posterior expectation of $\Pi_x\lVert\mathbf{x}_t-(W\mathbf{z}+\mathbf{b})\rVert^2-D_x\ln\Pi_x$, the precision update of variational Laplace \citep{Friston2007Laplace}. When clipping changes the mean, the update approximates the posterior expected reconstruction error. Eq.~\eqref{eq:supp:l2} omits $-D_x\ln\Pi_x$ because it does not affect inference over $\mathbf{z}$. The variance starts at the mean squared prediction error of the first step.

\paragraph{Learned access prior.}
L3 keeps Dirichlet counts $\boldsymbol{\alpha}^{\mathrm{acc}}$ over the four first-order thoughtseeds, initialized uniformly to $\alpha_0$, with $\boldsymbol{\beta}=\boldsymbol{\alpha}^{\mathrm{acc}}/\mathbf{1}^\top\boldsymbol{\alpha}^{\mathrm{acc}}\in[0,1]^4$. Only the count of \texttt{breath\_sensation} is updated. It gains $\max(0,m_{t_0}-m_{\mathrm{rest}})$ when \texttt{breath\_sensation} enters the workspace at step $t_0$ ($o_{t_0}=\texttt{breath\_sensation}\neq o_{t_0-1}$) and then holds it without interruption for $\tau_{\mathrm{dyn}}/dt$ steps. Clarity $m_{t_0}$ and the resting clarity $m_{\mathrm{rest}}$ (Eq.~\eqref{eq:supp:tonic}) are both taken at the re-entry step $t_0$, and the credit is added once the hold is complete.

\paragraph{Learned access counts and tonic contrast.}
L3 keeps access counts $\boldsymbol{\alpha}^{\mathrm{ton}}$ over the four first-order thoughtseeds, also initialized uniformly to $\alpha_0$. At each training step the count of the content holding the workspace increases by one, and an empty slot adds none. With $p_i=\alpha^{\mathrm{ton}}_i/\sum_j\alpha^{\mathrm{ton}}_j$, the tonic contrast applies the transform that gives $c(i)$ from $1-d_i(\mathrm{MW})$ to the count-weighted mean of $1-d_i(\mathrm{MW})$,
\begin{equation}
g = \ln\frac{\pi}{1-\pi} - \ln 3, \qquad \pi = \sum_i p_i\,\bigl[1-d_i(\mathrm{MW})\bigr].
\label{eq:supp:tonic}
\end{equation}
Since $b_m$ is the value of $g$ for uniform counts (Table~\ref{tab:core_params}), $m_{\mathrm{rest}}=\tfrac12$ before training, and the learned counts set where the monitor rests. Both phenotypes use the same rule, initial counts and training exposure, so differences in $\boldsymbol{\beta}$ and $g$ arise through their trajectories under the configured dynamics.

\paragraph{Learned return habit.}
L3 keeps habit counts $\boldsymbol{\alpha}^{\mathrm{hab}}$ over the two MA successors, returning (RA) and not returning (MW), with $E=\boldsymbol{\alpha}^{\mathrm{hab}}/\mathbf{1}^\top\boldsymbol{\alpha}^{\mathrm{hab}}$. The prior counts are the phenotype's MA transition row times $\alpha_0$ (Table~\ref{tab:core_params}). At each training step with a detection event ($\Delta_t=1$), the return act $P(\pi)$ of that step is added to $\boldsymbol{\alpha}^{\mathrm{hab}}$, one count per detection event, as in active-inference habit learning from the policies an agent forms \citep{Friston2016Learning}. Because the configured scores favor RA, these updates raise the habit's probability of returning.

\section{Parameters}

Numerical settings are configured, derived or learned within the simulation, and none is fitted to empirical data.

\begin{table}[t]
\centering
\caption{\textbf{Shared settings and learned values.} Learned values are from the illustrative run (training seed \StatsSeed{}) after training and are frozen during the rollout. Configured settings are modeling choices, and derived ones are computed from other settings or tables. Times are in seconds unless given in steps.}
\label{tab:core_params}
\small
\begin{tabular}{>{\raggedright\arraybackslash}p{5.4cm}>{\raggedright\arraybackslash}p{3.8cm}>{\raggedright\arraybackslash}p{4.8cm}}
\hline
Parameter & Value & Kind \\
\hline
\multicolumn{3}{l}{\textit{Simulation and timescales}}\\
Time step $dt$ & 0.2 & Configured \\
Training phase & \NovicePracticeSteps{} steps per phenotype & Configured \\
Frozen rollout & \StatsEvaluationSteps{} steps, evaluation window the final \StatsPlotSteps{} & Configured \\
Layer timescale separation $k_{\mathrm{sep}}$ & 2.0 & Configured, sets the L1 settling-margin bound $k_{\mathrm{sep}}/\tau_{\mathrm{dyn}}$ and $\tau_m=k_{\mathrm{sep}}\tau_{\mathrm{dyn}}$ \\
L2 time constant $\tau_{\mathrm{dyn}}$ & 1.0 & Configured \\
Monitoring-clarity time constant $\tau_m$ & 2.0 & Derived, $k_{\mathrm{sep}}\tau_{\mathrm{dyn}}$ \\
\multicolumn{3}{l}{\textit{L1 and L2}}\\
Diffusion variance rate $\sigma_x^2$ & 0.002 & Configured \\
Dwell coefficient of variation $c_v$ & 0.6 & Configured \\
Learning rate $\eta$, decoder and observation precision & 0.01 & Configured \\
Observation precision $\Pi_x$ & \ExpertObservationPrecision{} (expert), \NoviceObservationPrecision{} (novice) & Learned (Eq.~\eqref{eq:supp:obs_precision}) \\
\multicolumn{3}{l}{\textit{L3 access and monitoring}}\\
Access threshold & 0.5 & Configured, low enough that distracting content can stay in the workspace while the lapse is recognized \\
Access reverberation $\kappa$ & 0.25 & Configured \\
Access adaptation weight $w_a$ & 1.15 & Derived, $z_{\max}+\kappa$ \\
Access adaptation time constant $\tau_a$ & 5.0 & Configured \\
Initial concentration $\alpha_0$ & 20 & Configured, for the access prior, the access counts and the habit \\
Access-credit hold & 5 steps (1\,s) & Derived, $\tau_{\mathrm{dyn}}/dt$ \\
Breath access prior $\beta_{\mathrm{b}}$ & \ExpertBreathAccessPrior{} (expert), \NoviceBreathAccessPrior{} (novice) & Learned (\S S1.6) \\
Monitor bias $b_m$ & $+0.04$ & Derived, $g$ for uniform access counts \\
Recognition-cue monitor term & gain $c(\mathrm{rc})=1.02$, centre $\mu_{z,\mathrm{rc}}(\mathrm{MW})=0.15$ & Derived from Table~\ref{tab:thoughtseed_priors} \\
Monitoring-check probability $p_k$ & 0.0106 per distractor step & Configured, calibrated for illustration (\S S1.4) \\
Check interval & 10 steps (2\,s) & Derived, $\tau_m/dt$ \\
Check contrast $c_k$ & 0.715 & Derived, $c(\mathrm{rc})[\mu_{z,\mathrm{rc}}(\mathrm{MA})-\mu_{z,\mathrm{rc}}(\mathrm{MW})]$ \\
Resting clarity $m_{\mathrm{rest}}=\sigma(g-b_m)$ & \ExpertRestingClarity{} (expert), \NoviceRestingClarity{} (novice) & From learned counts (Eq.~\eqref{eq:supp:tonic}) \\
\multicolumn{3}{l}{\textit{Return policy}}\\
Scores $G(\mathrm{RA})$, $G(\mathrm{MA})$, $G(\mathrm{MW})$ & 1.20, 2.32, 2.32 & Derived (Eqs.~\eqref{eq:supp:policy} and~\eqref{eq:supp:policy_exit}) \\
Habit $E(\mathrm{RA})$ & \ExpertLearnedHabitReturn{} (expert), \NoviceLearnedHabitReturn{} (novice), from 0.95 and 0.70 & Learned (\S S1.6), prior counts the MA row of Table~\ref{tab:dwell_transitions} times $\alpha_0$ \\
\hline
\end{tabular}
\end{table}

\begin{table}[t]
\centering
\caption{\textbf{Configured dwell means and transition probabilities by phenotype, with realized mean dwells.} The configured mean $T_s$ and the shared coefficient of variation $c_v=0.6$ define each regime's baseline dwell-time distribution. The realized column is the mean over fully observed episodes in the evaluation window of the illustrative run, and Table~\ref{tab:robustness} gives paired expert-minus-novice dwell differences across five runs. Realized durations also depend on broadcast stabilization and, in MA, on the return policy, so $T_s$ is not an expected realized mean. The successor columns give the configured $P(s'\mid s)$, with self-transitions omitted (--). Outside MA, L1 draws successors from these rows. The MA row initializes the learned habit $E$ (\S S1.6), and MA's successors are drawn from the return policy $P(\pi)$, which favors returning more than $E$ when $\gamma_t>0$ (Eq.~\eqref{eq:supp:policy}).}
\begin{tabular}{llcccccc}
\hline
 & & \multicolumn{2}{c}{Mean dwell (s)} & \multicolumn{4}{c}{Successor $s'$}\\
Phenotype & Regime $s$ & $T_s$ & Realized & BF & MW & MA & RA \\
\hline
Expert & BF & 20.0 & \ExpertDwellBreathFocus{} & -- & 0.60 & 0.20 & 0.20 \\
Expert & MW & 14.0 & \ExpertDwellMindWandering{} & 0.05 & -- & 0.80 & 0.15 \\
Expert & MA & 4.5 & \ExpertDwellMetaAwareness{} & 0.00 & 0.05 & -- & 0.95 \\
Expert & RA & 3.0 & \ExpertDwellRedirectAttention{} & 0.70 & 0.10 & 0.20 & -- \\
Novice & BF & 14.0 & \NoviceDwellBreathFocus{} & -- & 0.80 & 0.10 & 0.10 \\
Novice & MW & 20.0 & \NoviceDwellMindWandering{} & 0.10 & -- & 0.75 & 0.15 \\
Novice & MA & 7.5 & \NoviceDwellMetaAwareness{} & 0.00 & 0.30 & -- & 0.70 \\
Novice & RA & 5.0 & \NoviceDwellRedirectAttention{} & 0.60 & 0.25 & 0.15 & -- \\
\hline
\end{tabular}
\label{tab:dwell_transitions}
\end{table}

\begin{table*}[t]
\centering
\caption{\textbf{Thoughtseed prior means $\boldsymbol{\mu}_z(s)$, also the L2 attractors.} Configured and shared by both phenotypes. The lower block gives the content quantities computed from these entries (\S S1.4), rounded to two decimals. The preference $C$ is normalized over the four first-order contents, and dashes mark quantities not used for \texttt{recognition\_cue}, a monitoring thoughtseed, which does not hold the workspace. MA has the MW priors for the four first-order thoughtseeds and a higher \texttt{recognition\_cue} prior.}
\footnotesize
\setlength{\tabcolsep}{4pt}
\begin{tabular}{lccccc}
\hline
Regime & \begin{tabular}[b]{@{}c@{}}breath\_\\sensation\end{tabular} & \begin{tabular}[b]{@{}c@{}}pain\_\\discomfort\end{tabular} & \begin{tabular}[b]{@{}c@{}}pending\_\\tasks\end{tabular} & \begin{tabular}[b]{@{}c@{}}recognition\_\\cue\end{tabular} & \begin{tabular}[b]{@{}c@{}}intention\_to\_\\return\end{tabular} \\
\hline
Breath Focus (BF) & 0.85 & 0.20 & 0.05 & 0.15 & 0.45 \\
Mind Wandering (MW) & 0.15 & 0.65 & 0.60 & 0.15 & 0.10 \\
Meta Awareness (MA) & 0.15 & 0.65 & 0.60 & 0.85 & 0.10 \\
Redirect Attention (RA) & 0.70 & 0.15 & 0.15 & 0.25 & 0.85 \\
\hline
\multicolumn{6}{l}{\textit{Computed from the priors above (\S S1.4)}}\\
Diagnosticity $d_i(\mathrm{BF})$ & 0.46 & 0.12 & 0.04 & 0.11 & 0.30 \\
Diagnosticity $d_i(\mathrm{MW})$ & 0.08 & 0.39 & 0.43 & 0.11 & 0.07 \\
Diagnosticity $d_i(\mathrm{MA})$ & 0.08 & 0.39 & 0.43 & 0.61 & 0.07 \\
On-task contrast $c(i)$ & $+1.33$ & $-0.67$ & $-0.81$ & $+1.02$ & $+1.54$ \\
On-task weight $w_i$ & $+0.58$ & $-0.32$ & $-0.38$ & -- & $+0.65$ \\
Lapse evidence $\lambda(i)$ & $-0.94$ & $+0.88$ & $+2.20$ & -- & $-0.25$ \\
Preference $C(i)$ & 0.50 & 0.13 & 0.04 & -- & 0.33 \\
\hline
\end{tabular}
\label{tab:thoughtseed_priors}
\end{table*}

\FloatBarrier

\section{Outcome Measures}

Unless stated otherwise, summaries use the evaluation window of the main text, the final \StatsPlotSteps{} frozen-rollout steps. A regime run, a workspace-content run or a recognition interval is \emph{left-cut} if it began before the window and \emph{open} if it is still running at the final step. Durations are numbers of steps times $dt$. Detection events, the recognition interval, its overlap with the same distractor, breath re-entry into the workspace and the regime return from MA to RA are measured separately and need not coincide.

\paragraph{Regime and access measures.}
\begin{itemize}
\setlength{\itemsep}{0pt}
\item \textbf{Regime dwell.} Means over episodes that are neither left-cut nor open.
\item \textbf{Transitions.} Exits within the window to each destination, divided by the source regime's exit count in the window, with self-transitions excluded.
\item \textbf{Workspace access.} Entries of a content into the slot within the window, so a left-cut run is not an entry. Mean breath-access run length averages the portions of breath-access runs observed within the window, including left-cut and open runs, unlike regime dwell.
\item \textbf{Occupancy and monitoring.} Breath-access percentage is the percentage of window steps with \texttt{breath\_sensation} in the slot. Distractor access during MW is the fraction of MW steps occupied by either distractor. Mean BF monitoring averages $m_t$ over BF steps.
\item \textbf{Return effort.} Defined in Section~2.3 of the main text and averaged over MA exits within the window, left-cut MA episodes included. Training-block values use the habit that formed each applied policy, replayed from the saved detection events.
\end{itemize}

\paragraph{Recognition measures.} The recognition-anchored measures follow each detection event in the evaluation window, in any regime, to the recording end. Each event has one of four outcomes, the breath already in the workspace at the event, breath re-entry before or at recognition offset, recognition ending first, or the recognition interval remaining open at the window end without breath re-entry. All four enter the denominator of the breath re-entry proportion, and only the second enters its numerator and the recognition-to-breath latency. Recognition ends at the first step at which $\psi(m_t,q_t)$ fails, and the ending is attributed to the belief ($q_t<\tfrac12$), to clarity ($m_t\le m_{\mathrm{rest}}$), or to both when both hold at that step. Completed same-distractor overlaps are included even when the recognition interval is still open at the window end, and overlaps still open at the window end are excluded from duration summaries. Table~2 of the main text gives medians and ranges of per-run summaries, Table~\ref{tab:onset_outcomes} pools event counts over the five runs, and Table~\ref{tab:robustness} gives paired expert-minus-novice differences.

The proportion of distraction time with recognition counts steps rather than events. It is the share of window steps with a distractor in the workspace at which $D_t=1$, over all regimes and within Mind Wandering and Meta Awareness, and it is unavailable without such steps.

Recognition-to-RA latency uses MA episodes that begin in the window and contain a detection event (Section~2.3 of the main text). Detections outside MA do not enter it, and an MA episode entered with the lapse already recognized contains no event unless recognition ends and a new event occurs. Only observed direct exits to RA are timed, from the episode's first event.

Recognition during established Mind Wandering counts detection events in MW at least 3\,s after MW entry, with the entry taken from the full recording. Its rate divides their number by the time in established MW within the window. Episode measures use MW episodes that begin and end within the window and last longer than 3\,s. An episode counts as without a recognition onset after 3\,s when no detection event occurs after that point, so recognition carried across it without a new event does not count as an onset. Its longest stretch without recognition is the longest run of steps with $D_t=0$ over the whole episode. Five-run values are pooled over the windows.

\section{Supplementary Results}

\begin{figure}[t]
\centering
\begin{tabular}{c}
\includegraphics[width=0.66\textwidth]{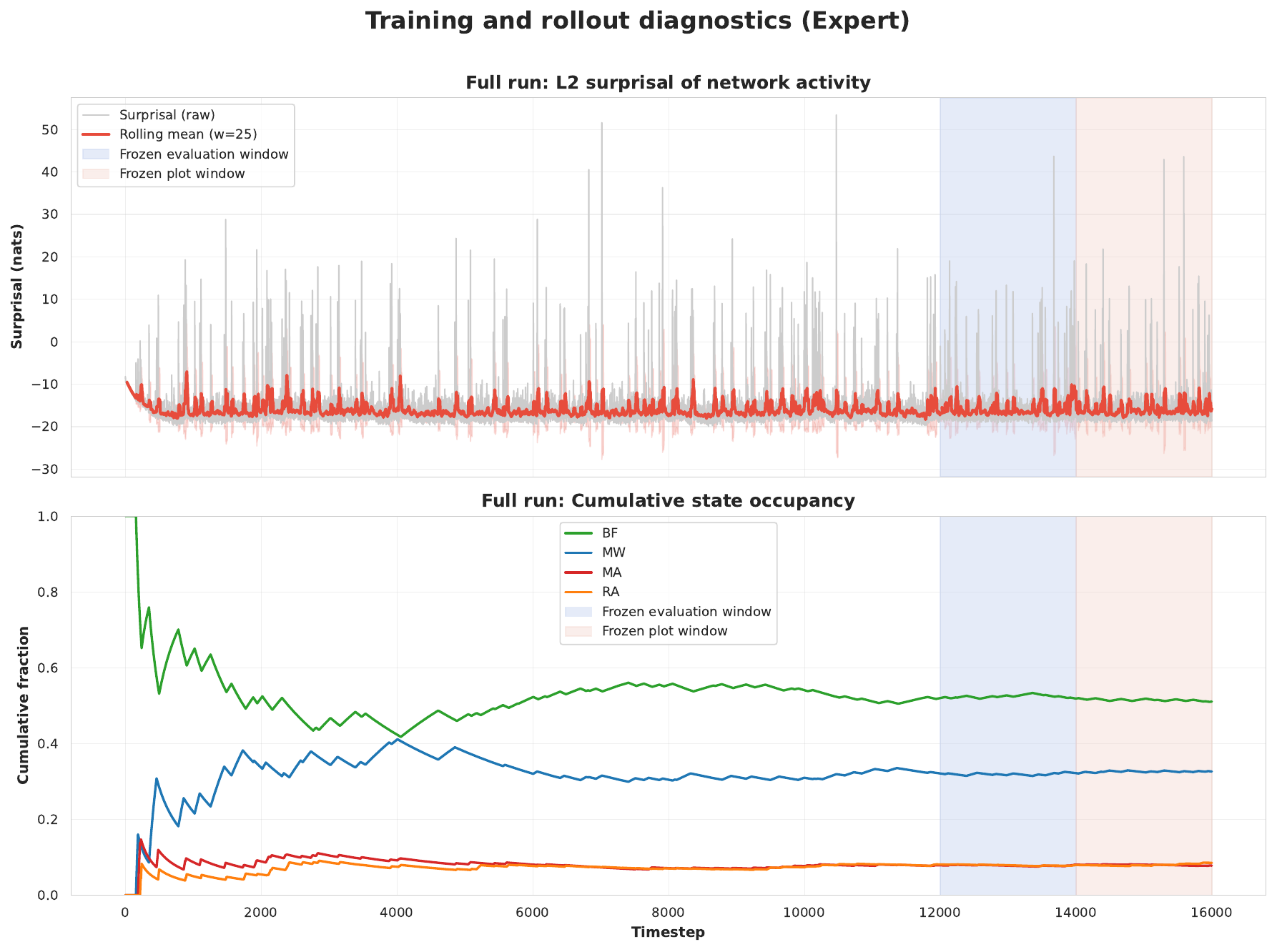}\\
\small{(A) Expert. Training, frozen evaluation and frozen plot window}
\end{tabular}
\vspace{0.5em}

\begin{tabular}{c}
\includegraphics[width=0.66\textwidth]{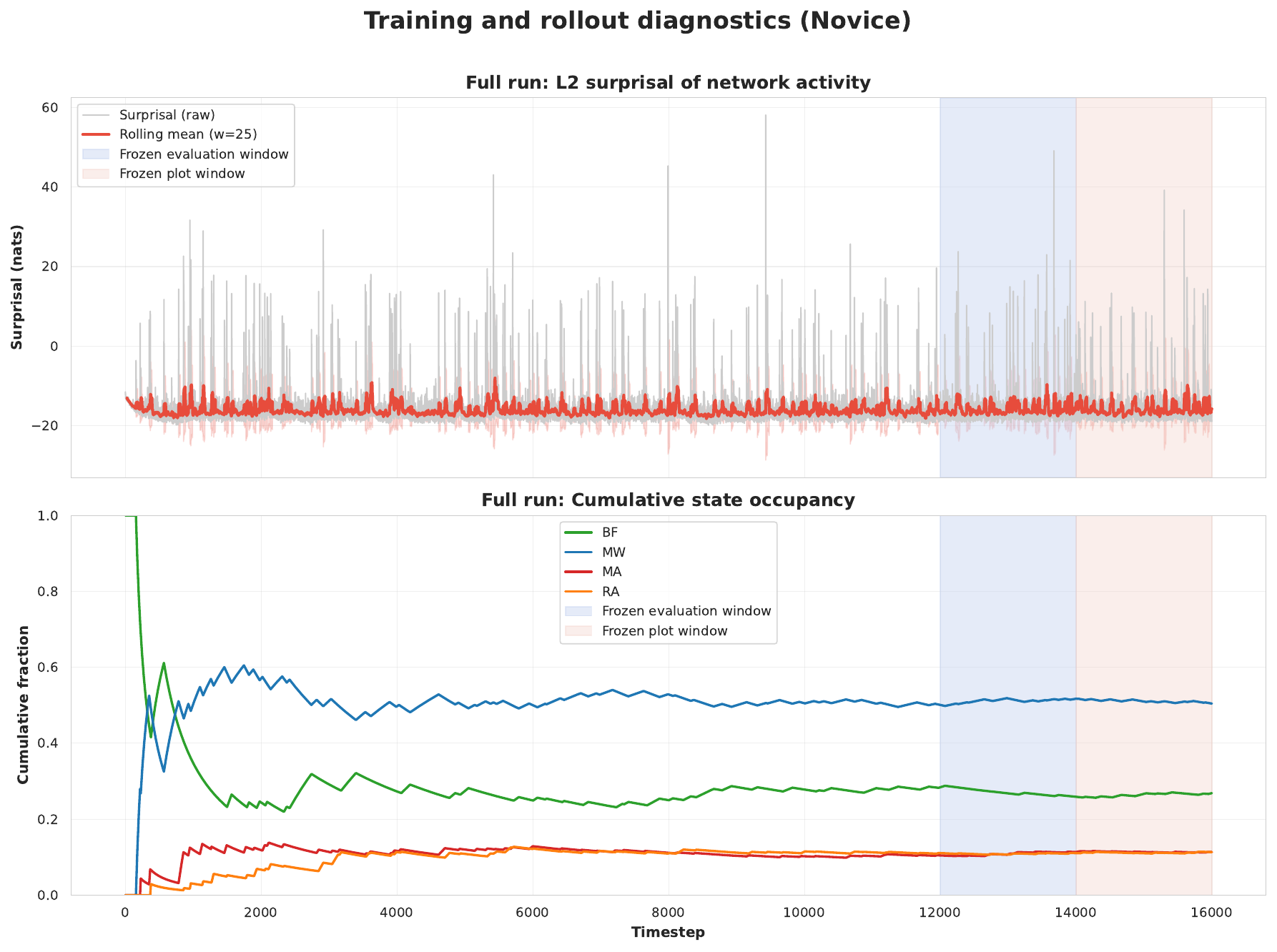}\\
\small{(B) Novice. Training, frozen evaluation and frozen plot window}
\end{tabular}
\caption{\textbf{Training, evaluation and plot diagnostics.} L2 surprisal of network activity and cumulative regime occupancy during training (\ExpertPracticeSteps{} expert, \NovicePracticeSteps{} novice steps) and an independent \StatsEvaluationSteps{}-step frozen rollout, expert in (A) and novice in (B). Panels show one illustrative training run and its frozen rollout per phenotype, with training seed \StatsSeed{}. The surprisal is the negative log predictive density of the network activity under L2's model (\S S1.3), in nats, and is negative where the density exceeds one. It falls early in training and then fluctuates around a stable level, with its largest values at regime switches. Blue marks the first \StatsInitialWindowSteps{} rollout steps and orange the final \StatsPlotSteps{} steps used by Figs.~4--6 of the main text. The trace break marks the start of the frozen rollout.}
\label{fig:S1}
\par\smallskip\noindent{\small Alt text: Two panels, expert above and novice below. Each shows the L2 surprisal of network activity and cumulative regime occupancy across training and the frozen rollout, with a trace break at the rollout start, a blue span for the first evaluation steps and an orange span for the plot window.}
\end{figure}

\begin{figure}[t]
\centering
\includegraphics[width=0.95\textwidth]{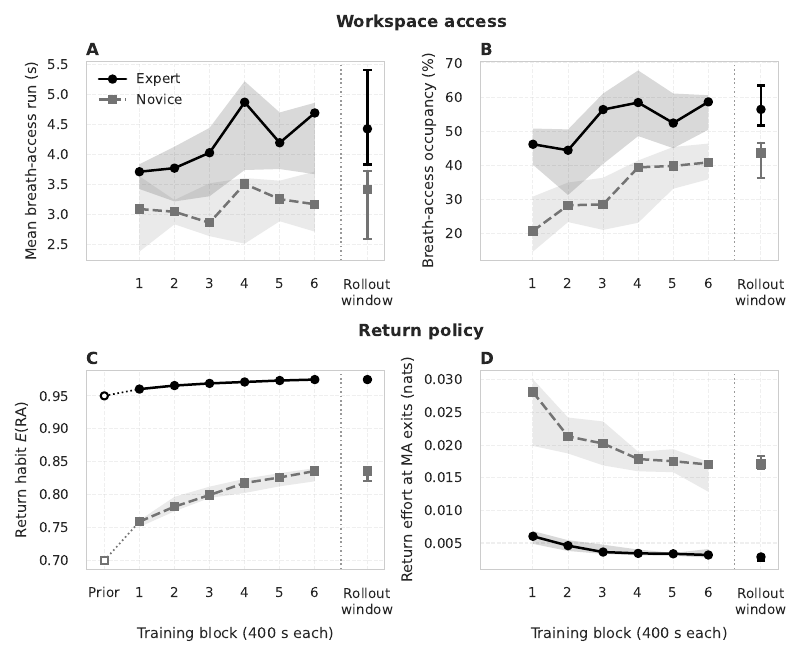}
\caption{\textbf{Practice curves over the five runs.} (A) Mean breath-access run length and (B) breath-access occupancy within each block, with runs cut by the block boundaries included. (C) The return habit's probability of returning $E(\mathrm{RA})$ at the end of each block, with the configured priors as open markers left of block 1. (D) Mean return effort at the block's MA exits (\S S3). The blocks are the six training blocks of \StatsPlotSteps{} steps, followed by the frozen-rollout window right of the dotted line. Lines and markers are medians across training seeds 100--104, and bands and whiskers are the range across runs. Solid black is the expert and dashed grey the novice. The configured dwell-time distributions stay fixed during training, and the mechanisms behind the lengthening of the expert's breath-access runs were not isolated.}
\label{fig:S2}
\par\smallskip\noindent{\small Alt text: Four panels of lines over six training blocks with a separate rollout point at the right, workspace access in the upper row and the return policy in the lower row. Breath-access occupancy rises over training for both phenotypes, breath-access run length rises clearly only for the expert, and the expert's values stay above the novice's. The return habit rises slowly from its configured prior, marked by an open symbol, for both phenotypes and the expert's is higher. Return effort falls for both and the expert's is lower.}
\end{figure}

Supplementary Fig.~\ref{fig:S1} gives training and rollout diagnostics, Tables~\ref{tab:robustness} and~\ref{tab:onset_outcomes} outcomes across the five runs, and Supplementary Fig.~\ref{fig:S2} practice-related changes over training. The illustrative Figs.~4--6 of the main text use training seed \StatsSeed{}, chosen by inspection of the five runs as the one whose plot window shows the lapse, recognition and return sequence of both phenotypes most clearly. It is an illustration, not a representative run, and its differences need not follow every median direction. The five-run summaries describe stochastic variability, not hypothesis tests or population-level effect sizes.

\input{figures/robustness.tex}
\ifdefined\RobustnessModelVersion
  \edef\ActualRobustnessVersion{\RobustnessModelVersion}
  \ifx\ActualRobustnessVersion\ExpectedModelVersion\else
    \errmessage{Robustness table uses an older model: rerun scripts/aggregate_robustness.py}
  \fi
\else
  \errmessage{Robustness table predates MA detection: rerun all seeds and scripts/aggregate_robustness.py}
\fi

\paragraph{Recognition offset after breath re-entry.} Over the five runs, \FiveRunExpertReentryOffsetEligible{} expert and \FiveRunNoviceReentryOffsetEligible{} novice recognition onsets had breath re-entry before or at recognition offset, and the offset was observed by the window end in \FiveRunExpertReentryOffsetTimed{} and \FiveRunNoviceReentryOffsetTimed{} of them. The median of per-run median times from the first breath re-entry to recognition offset was \FiveRunExpertReentryOffsetMedian{}~s (\FiveRunExpertReentryOffsetRange{}) in the expert and \FiveRunNoviceReentryOffsetMedian{}~s (\FiveRunNoviceReentryOffsetRange{}) in the novice, so recognition usually continued beyond re-entry. A distractor recurred in the workspace between the first breath re-entry and recognition offset in \FiveRunExpertReentryOffsetRecurred{} of \FiveRunExpertReentryOffsetTimed{} expert and \FiveRunNoviceReentryOffsetRecurred{} of \FiveRunNoviceReentryOffsetTimed{} novice completed intervals.

\paragraph{Recognition during established Mind Wandering.} Pooled over the five evaluation windows, \FiveRunExpertChecks{} expert and \FiveRunNoviceChecks{} novice monitoring checks occurred. There were \FiveRunExpertEstablishedMWOnsets{} expert and \FiveRunNoviceEstablishedMWOnsets{} novice recognition onsets during established Mind Wandering, present in \FiveRunExpertEstablishedMWOnsetRuns{} and \FiveRunNoviceEstablishedMWOnsetRuns{} runs. The episode measures use \FiveRunExpertEstablishedMWEpisodes{} expert and \FiveRunNoviceEstablishedMWEpisodes{} novice complete episodes (Section~S3).

\paragraph{Monitoring-check probability.} As a sensitivity check, workspace access and monitoring were replayed open-loop over the saved rollouts of the preceding model version, which was trained and run without monitoring checks, with the regime sequence held fixed. A phenotype-specific probability equal to $\varepsilon_{\mathrm{off}}$ of Eq.~\eqref{eq:supp:belief_predict} replaced the shared $p_k$. In this replay, recognition during established Mind Wandering occurred in every run, most established episodes still had no recognition onset after 3\,s, and the share of onsets with a distractor in the workspace changed little. The coupled probability produced fewer onsets per minute of established Mind Wandering in the novice than in the expert, whereas the shared probability gave similar rates. The coupling imposes the phenotype ordering of the check probability itself. Because the regimes were fixed, the replay does not show effects on regime exits or learning.

\FloatBarrier

\section{Contemplative Foundations of the Architecture}
\label{sec:contemplative}

\subsection{Scope and Interpretation}

The practice setting is breath-directed focused-attention (FA) meditation,
motivated by the first tetrad of MN 118. The four-phase cycle of the main text
(Fig.~1) supplies the simulation's organization, and the three layers and their
update rules are computational choices. The model represents a restricted
lapse-and-recognition process within this setting, rather than the full
sixteen-step practice or its soteriological aims.

Early Buddhist discourses, Therav\=ada commentarial accounts, modern practice
descriptions and Yog\=ac\=ara seed imagery are treated as distinct sources rather
than a single doctrinal account, and J.~Krishnamurti's account of attention is a
further source outside these traditions. We distinguish \emph{motivating} correspondences, which informed a design
choice, from \emph{descriptive} analogies, which help interpret a model element.
These labels state the role of a source in our account, not the strength of
empirical evidence. Neither category makes a computational rule a consequence
of the source. Buddhist terms retain broader meanings than the modeled
variables, and the dynamics and timescales are specified in \S\S S1--S2.

\subsection{Contents, Qualities, Acts and Phases}

The architecture keeps four kinds of description apart. \emph{Contents} are
what is experienced, and at most one content holds the workspace. They are
the thoughtseeds. \emph{Qualities and conditions} describe how the mind is doing.
They are read from the dynamics and are not additional thoughtseeds. \emph{Acts} are what the practitioner
does. The only modeled act is the return policy with recognition-gated modulation,
which in MA weights exit timing and selects Redirect Attention or Mind Wandering (\S S1.4). \emph{Phases} are the four regimes, which set the
context for which contents are likely.

The qualities are readouts of quantities already in the model. Clarity is
monitoring clarity $m_t$, and regime stabilization is the broadcast
stabilization (Eq.~7 of the main text). Distraction is a distractor holding the workspace,
the experienced distraction of the main text. Dullness has no separate
variable, although breath access with $m_{t-1}<m_{\mathrm{rest}}$ and an empty workspace
can both occur. Effort is read out only as return effort at MA exits
(Table~\ref{tab:phenomenology}), and equanimity, which practice descriptions also name,
is not represented.

The breath appears at all three layers. In L1, somatomotor (SOM) network
activity is a network-level representation associated with the practice, and
its configured attractor is highest in Breath Focus. The model does not
simulate respiration. In L2, the \texttt{breath\_sensation} latent has a nonzero prior
in every regime, so breath content is represented continuously. In L3, breath
content holds the workspace only intermittently. Only workspace access
corresponds to the breath as experienced.

\begin{longtable}{>{\small\raggedright\arraybackslash}p{0.22\textwidth} >{\small\raggedright\arraybackslash}p{0.28\textwidth} >{\small\raggedright\arraybackslash}p{0.42\textwidth}}
\caption{\textbf{Phenomenological interpretation of model quantities.} How each quantity is interpreted, and the limit of that interpretation. Definitions are in Table~1 of the main text, and Table~\ref{tab:contemplative} lists the contemplative sources.}
\label{tab:phenomenology}\\
\hline
Model quantity & Interpreted as & Limit of the interpretation \\
\hline
\endfirsthead
\hline
Model quantity & Interpreted as & Limit of the interpretation \\
\hline
\endhead
Workspace content $o_t$ & The currently experienced content & Access is functional. Report, vividness and memory of the content are not modeled. An empty slot means that no first-order content is accessed. \\
Distractor holding the workspace & Experienced distraction & It does not by itself start recognition (Eq.~5 of the main text). \\
Activations $\mathbf{z}_t^{*}$ of contents not in the workspace & Inferred latent activity without workspace access & The activations shape access and monitoring, but their magnitudes are not measures of reported salience. \\
\texttt{recognition\_cue} activation & Recognition-related latent activity & Does not enter the workspace and is not a report of recognition. It contributes positively to monitoring drive when its activation exceeds its MW prior. \\
Monitoring clarity $m_t$ & Graded, task-aligned clarity of monitoring & An operational signal driven by latent content contrast and the recognition-cue term, responding with a delay. Experiential vividness is not modeled, and its magnitude does not itself measure whether a lapse is recognized. \\
Monitoring check and check interval $e_t$ & An unscheduled occasion on which monitoring re-engages during experienced distraction & A configured stochastic event with a shared probability. The model specifies when checks are eligible but not why a particular check occurs. A check is not a report or a deliberate act, and whether recognition follows depends on clarity and the lapse belief. \\
Lapse belief $q_t$ & The estimate that sustained breath focus is not established & Its target is approximated by the Breath Focus regime and its transition probabilities by dwell means. \\
Meta Awareness regime & Configured interval that favors conditions for recognition & Recognition itself is the detection event. L3 does not observe the regime. \\
Detection event $\Delta_t$ & Recognition of the lapse & A discrete operationalization that can occur outside Meta Awareness. An explicit experiential report is not modeled, and its correspondence with recognition in experience remains to be validated. \\
Recognition indicator $D_t$ & The lapse remaining recognized & The duration of the recognition interval is not validated against experience, and a nonzero duration does not establish sustained awareness. \\
\texttt{intention\_to\_return} in the workspace & Experienced intention to return & Its lapse evidence is weakly negative (Table~\ref{tab:thoughtseed_priors}). Its access does not reset the recognition indicator, but its evidence can contribute to ending recognition. It is distinct from the return policy. \\
Return policy $P(\pi)$, $\Lambda_t$ & Probabilities of leaving MA and, on leaving, of returning or continuing to wander & Modulation requires recognition at the preceding step, and otherwise the habit applies. Its forecast is a proxy built from configured priors, and staying is scored for one step. It does not hold the workspace and is not the completed act of returning. \\
Return effort & Departure of the successor distribution from the learned habit & A readout that does not act on the dynamics. Its training trajectory depends on the habit and on the policy precision. It is zero at MA exits for which $D_{t-1}=0$. It excludes the change in exit timing, and effort in sustaining Breath Focus is not represented. \\
Access prior $\beta$ & Practice-acquired readiness of the breath to be accessed & Credit goes to the breath sensation alone, and other learned access biases are outside the model. \\
\hline
\end{longtable}

\subsection{The Cycle Phase by Phase}

The paragraphs below follow the four regimes, which are the model's organizing
categories. Grouping contemplative sources under them is an interpretive
presentation. It does not claim that these sources prescribe the four-phase
computational cycle. Table~\ref{tab:contemplative} separates source concepts from their roles in the
model, and the regime priors of the contents are in
Table~\ref{tab:thoughtseed_priors}.

\begin{longtable}{>{\small\raggedright\arraybackslash}p{0.25\textwidth} >{\small\raggedright\arraybackslash}p{0.26\textwidth} >{\small\raggedright\arraybackslash}p{0.19\textwidth} >{\small\raggedright\arraybackslash}p{0.14\textwidth}}
\caption{\textbf{Correspondences between model elements and contemplative
sources.} Roles, motivating or descriptive, are defined in \S S5.1. Sutta references follow the \emph{Majjhima
Nik\=aya} numbering \citep{NanamoliBodhi1995}, commentarial references are to the
\emph{Visuddhimagga} \citep{Buddhaghosa1991}, and practice-oriented readings follow
\citet{Analayo2018}. Bengio's consciousness prior is included as a
complementary computational parallel.}
\label{tab:contemplative}\\
\hline
Model element & Term & Source & Role \\
\hline
\endfirsthead
\hline
Model element & Term & Source & Role \\
\hline
\endhead
MA and detection $\Delta_t$ & knowing a distracted mind as distracted & MN 10, contemplation of mind & Motivating \\
Mind Wandering regime & \emph{vikkhitta} (distracted) & MN 10 & Descriptive \\
Monitor $m_t$ & \emph{sampaja\~n\~na} (clear comprehension) & MN 10 & Motivating \\
Redirect Attention regime & attending to another, skillful theme & MN 20 & Motivating \\
Single-slot workspace & each \emph{citta} (moment of consciousness) takes a single \emph{\=aramma\d{n}a} (object) & \citet{Bodhi2000} & Descriptive \\
Serial access by repeated selection & seriality of cognitive moments & \emph{Visuddhimagga} XIV.115--118 & Motivating \\
Access prior $\beta_i$ & cultivation through repeated practice & MN 118 & Motivating \\
Conditioning of activation and access & \emph{sa\d{m}sk\=ara} (Pali \emph{sa\.{n}kh\=ara}), volitional formations, intentional activity conditioning later experience & \citep{Bodhi1999} & Motivating \\
Selective workspace access & consciousness prior, a small conscious state selected from a richer representation & \citep{Bengio2017} & Descriptive \\
Dormant inventory & \emph{anusaya} (underlying tendencies) & \citet{Analayo2018} & Descriptive \\
\texttt{breath\_sensation} & \emph{\=an\=ap\=ana}, the breath as \emph{\=aramma\d{n}a} (practice object) & MN 118 and \citep[chap.~6]{Hart1988} & Descriptive \\
\texttt{pain\_discomfort} & \emph{dukkha vedan\=a} (unpleasant feeling), itself an object of \emph{vedan\=anupassan\=a} & MN 10 & Descriptive \\
\texttt{pending\_tasks} & loose analogy with \emph{uddhacca-kukkucca} (restlessness and worry), one of the five hindrances & MN 10, hindrances & Descriptive \\
\texttt{recognition\_cue} & monitoring thoughtseed contributing to monitoring clarity, which does not hold the workspace & MN 10 (see MA and detection) & Descriptive \\
Detection $\Delta_t$ and recognition indicator $D_t$ & awareness of inattention & \citet{Krishnamurti1969} & Descriptive \\
\texttt{intention\_to\_return} & returning attention to the breath, as in breath-meditation instructions & \citep[chap.~6]{Hart1988} and \citet{Analayo2018} & Descriptive \\
Thoughtseed as seed & \emph{b\={\i}ja} (seed) in \emph{\=alaya-vij\~n\=ana} & \citep{Waldron2003} & Descriptive \\
Specified conditioning relations & dependent arising (\emph{prat\={\i}tya\allowbreak samutp\=ada}) & \citep{Analayo2021} & Descriptive \\
Sustained breath access & continuity of attention to the breath & \emph{Visuddhimagga} VIII.195--196 & Descriptive \\
\hline
\end{longtable}

\paragraph{Breath Focus.} In the early texts the breath is the object (\emph{\=aramma\d{n}a}) of the first tetrad of MN 118, and practice descriptions place attention on the natural sensation of the in- and out-breath and return it there repeatedly (\citealp[chap.~6]{Hart1988}; \citealp{Analayo2018}). Commentarial accounts of cognitive moments and their objects (\citealp[XIV.115--118]{Buddhaghosa1991}) motivate serial access, which the single-slot workspace realizes as repeated selection of the same content at a numerically chosen update interval. Descriptions of sustained breath attention (\citealp[VIII.195--196]{Buddhaghosa1991}) offer a descriptive parallel. MN 118 also emphasizes developing and cultivating mindfulness of breathing. The learned access prior is our proposed representation of this practice-dependent tendency. Its breath-only credit rule (\S S1.6) privileges the practice object, so practice also lowers the access prior of \texttt{intention\_to\_return}. This motivates the hypothesis, untested here, that cultivation may shift experience from repeated focal intentions to more direct engagement with the breath.

\paragraph{Mind Wandering.} The Mind Wandering regime corresponds descriptively to the distracted mind (\emph{vikkhitta}, MN 10). Its two distractor contents are distractors only relative to the breath task. \texttt{pain\_discomfort} is the unpleasant bodily sensation that arises as sitting continues. Because unpleasant feeling (\emph{dukkha vedan\=a}) is itself an object of contemplation (\emph{vedan\=anupassan\=a}, MN 10), its distractor status holds only because it occupies the slot instead of breath content. \texttt{pending\_tasks} is the planning of unfinished obligations. It is content, whereas restlessness and worry (\emph{uddhacca-kukkucca}) is a quality of mind, so the correspondence with that hindrance (MN 10) is a loose descriptive analogy.

\paragraph{Meta Awareness and recognition.} Knowing a distracted mind as distracted (MN 10) motivates the distinction between experiencing a distraction and recognizing it as a lapse. The verb \emph{paj\=an\=ati} describes knowing or discerning, and its use here motivates recognition without specifying a discrete computational event. Detection (\S S1.4) is our implementation of recognition. Giving Meta Awareness the Mind Wandering priors for the first-order contents, with a raised \texttt{recognition\_cue}, lets recognition begin while the distractor remains in the workspace. This coexistence is a modeling choice, not a dynamical claim made by the source, which motivates only the distinction. The recognition indicator operationalizes maintained recognition, and its brief overlap with the same distractor (Table~2 of the main text) gives limited coverage of sustained recognition during ongoing distraction. Meta Awareness can end without detection. Clear comprehension (\emph{sampaja\~n\~na}, MN 10) is our conceptual motivation for monitoring clarity. Its graded, content-driven dynamics are a modeling choice, it can be high during sustained breath access with no lapse to detect, and identifying $m_t$ with an empirical measure of clear comprehension would require a separate measurement relation. J.~Krishnamurti's statement that the awareness of inattention is attention \citep{Krishnamurti1969} offers a descriptive parallel to recognition of the lapse. The parallel is partial, because Krishnamurti describes this awareness as choiceless and as attention itself, whereas the model keeps recognition distinct from the return, which a separate policy selects.

\paragraph{Redirect Attention and return.} The first method in MN 20 directs attention to another, skillful theme when unskillful thoughts arise, which motivates supporting alternative content. Our Redirect Attention regime supplies priors favoring breath content and the intention to return, while competition determines the workspace occupant. Its application to the breath task is a modeling analogy, and the source's ethical classification of thoughts is distinct from our on-task versus off-task distinction. Breath-meditation instructions describe returning attention to the breath (\citealp[chap.~6]{Hart1988}; \citealp{Analayo2018}). The content \texttt{intention\_to\_return} represents the experienced intention to return, our representation rather than a distinct latent intention identified in those sources. It is a first-order content that concerns attention itself, and it is distinct from the return policy and from the realized transition to Redirect Attention (Table~\ref{tab:phenomenology}). Reading the policy's recognition-gated modulation as the return that follows noticing a lapse is a bounded modeling interpretation, which the sources do not specify, and the habit route without recognition is likewise a modeling choice.

\subsection{Formations, Seeds and the Dormant Inventory}

\emph{Sa\d{m}sk\=ara} (Pali \emph{sa\.{n}kh\=ara}) denotes volitional formations,
intentional activity that conditions later experience \citep{Bodhi1999}. We read
meditation as learned in this way, through repeated volitional practice that
shapes what later arises and what is attended to. This reading motivates the
thoughtseed framework, which distinguishes a dormant repertoire of potential
contents, a context-active subset engaged in ongoing processing, and the content
selected for workspace access. In the implementation, the
practice-learning analogy concerns the access prior (\S S1.6). The
correspondence is interpretive, since formations motivate the conditioning of
activation and access, while thoughtseeds remain the model's content variables.
The simulation fixes the context-active repertoire, and
recruitment from the dormant inventory remains conceptual.

The dormant inventory offers a limited analogy with underlying tendencies
(\emph{anusaya}; \citealp{Analayo2018}). The limitation matters, because
\emph{anusaya} denotes unwholesome tendencies, whereas the proposed inventory
also includes cultivated dispositions. We retain ``dormant inventory'' as the
architectural term. The seed metaphor (\emph{b\={\i}ja} in
\emph{\=alaya-vij\~n\=ana}; \citealp{Waldron2003}) and the analogy with dependent
arising (\emph{prat\={\i}tyasamutp\=ada}; \citealp{Analayo2021}) supply imagery
for dispositions conditioning later experience and for specified conditioning
relations, not an implemented storehouse or a taxonomy of latent variables.

\subsection{Limits of Interpretation}

The empty workspace as failed access, and the release of broadcast
stabilization during recognition, are operational definitions. The simulation addresses functional access, monitoring and
regulation, whereas doctrinal accounts of life-continuum, liberation, absorption and
the detailed cognitive series lie outside its scope.

Specialist review may refine the interpretation of knowing distraction, the
relationship between clear comprehension (\emph{sampaja\~n\~na}) and
mindfulness (\emph{sati}), and the limited use of latent-tendency imagery.

\section{Supplementary References}

%% file: figures/stats.tex
\newcommand{\StatsModelVersion}{clarity-v48}
\newcommand{\StatsSeed}{103}
\newcommand{\StatsPlotSteps}{2,000}
\newcommand{\StatsEvaluationSteps}{4,000}
\newcommand{\StatsInitialWindowSteps}{2,000}
\newcommand{\StatsAnatomySeed}{103}
\newcommand{\ExpertDwellBreathFocus}{24.0}
\newcommand{\ExpertBreathFocusDMN}{0.348}
\newcommand{\ExpertBreathFocusVAN}{0.448}
\newcommand{\ExpertBreathFocusDAN}{0.652}
\newcommand{\ExpertBreathFocusFPN}{0.700}
\newcommand{\ExpertBreathFocusVIS}{0.251}
\newcommand{\ExpertBreathFocusSOM}{0.650}
\newcommand{\ExpertBreathFocusLIM}{0.249}
\newcommand{\ExpertTransitionBreathFocusToMindWandering}{0.88}
\newcommand{\ExpertTransitionBreathFocusToMetaAwareness}{0.00}
\newcommand{\ExpertTransitionBreathFocusToRedirectAttention}{0.12}
\newcommand{\ExpertDwellMindWandering}{16.0}
\newcommand{\ExpertMindWanderingDMN}{0.695}
\newcommand{\ExpertMindWanderingVAN}{0.400}
\newcommand{\ExpertMindWanderingDAN}{0.287}
\newcommand{\ExpertMindWanderingFPN}{0.386}
\newcommand{\ExpertMindWanderingVIS}{0.349}
\newcommand{\ExpertMindWanderingSOM}{0.288}
\newcommand{\ExpertMindWanderingLIM}{0.545}
\newcommand{\ExpertTransitionMindWanderingToBreathFocus}{0.00}
\newcommand{\ExpertTransitionMindWanderingToMetaAwareness}{0.78}
\newcommand{\ExpertTransitionMindWanderingToRedirectAttention}{0.22}
\newcommand{\ExpertDwellMetaAwareness}{3.5}
\newcommand{\ExpertMetaAwarenessDMN}{0.579}
\newcommand{\ExpertMetaAwarenessVAN}{0.818}
\newcommand{\ExpertMetaAwarenessDAN}{0.408}
\newcommand{\ExpertMetaAwarenessFPN}{0.599}
\newcommand{\ExpertMetaAwarenessVIS}{0.301}
\newcommand{\ExpertMetaAwarenessSOM}{0.382}
\newcommand{\ExpertMetaAwarenessLIM}{0.488}
\newcommand{\ExpertTransitionMetaAwarenessToBreathFocus}{0.00}
\newcommand{\ExpertTransitionMetaAwarenessToMindWandering}{0.00}
\newcommand{\ExpertTransitionMetaAwarenessToRedirectAttention}{1.00}
\newcommand{\ExpertDwellRedirectAttention}{5.0}
\newcommand{\ExpertRedirectAttentionDMN}{0.307}
\newcommand{\ExpertRedirectAttentionVAN}{0.424}
\newcommand{\ExpertRedirectAttentionDAN}{0.813}
\newcommand{\ExpertRedirectAttentionFPN}{0.716}
\newcommand{\ExpertRedirectAttentionVIS}{0.286}
\newcommand{\ExpertRedirectAttentionSOM}{0.595}
\newcommand{\ExpertRedirectAttentionLIM}{0.235}
\newcommand{\ExpertTransitionRedirectAttentionToBreathFocus}{0.80}
\newcommand{\ExpertTransitionRedirectAttentionToMindWandering}{0.20}
\newcommand{\ExpertTransitionRedirectAttentionToMetaAwareness}{0.00}
\newcommand{\ExpertReturnEffort}{0.003}
\newcommand{\ExpertLearnedHabitReturn}{0.975}
\newcommand{\ExpertReturnEffortFitFirst}{0.006}
\newcommand{\ExpertReturnEffortFitLast}{0.003}
\newcommand{\ExpertPracticeSteps}{12,000}
\newcommand{\ExpertAccessEvents}{149}
\newcommand{\ExpertBreathAccessPercent}{53.8}
\newcommand{\ExpertBreathRunSeconds}{4.3}
\newcommand{\ExpertDistractorAccessMWPercent}{80.9}
\newcommand{\ExpertClarityBF}{0.79}
\newcommand{\ExpertClaritySession}{0.68}
\newcommand{\ExpertBreathAccessPrior}{0.53}
\newcommand{\ExpertObservationPrecision}{1984}
\newcommand{\ExpertRestingClarity}{0.57}
\newcommand{\ExpertRegimeBeliefPercent}{99.9}
\newcommand{\NoviceDwellBreathFocus}{17.7}
\newcommand{\NoviceBreathFocusDMN}{0.495}
\newcommand{\NoviceBreathFocusVAN}{0.450}
\newcommand{\NoviceBreathFocusDAN}{0.583}
\newcommand{\NoviceBreathFocusFPN}{0.600}
\newcommand{\NoviceBreathFocusVIS}{0.304}
\newcommand{\NoviceBreathFocusSOM}{0.550}
\newcommand{\NoviceBreathFocusLIM}{0.350}
\newcommand{\NoviceTransitionBreathFocusToMindWandering}{0.86}
\newcommand{\NoviceTransitionBreathFocusToMetaAwareness}{0.14}
\newcommand{\NoviceTransitionBreathFocusToRedirectAttention}{0.00}
\newcommand{\NoviceDwellMindWandering}{18.4}
\newcommand{\NoviceMindWanderingDMN}{0.815}
\newcommand{\NoviceMindWanderingVAN}{0.351}
\newcommand{\NoviceMindWanderingDAN}{0.305}
\newcommand{\NoviceMindWanderingFPN}{0.336}
\newcommand{\NoviceMindWanderingVIS}{0.418}
\newcommand{\NoviceMindWanderingSOM}{0.303}
\newcommand{\NoviceMindWanderingLIM}{0.744}
\newcommand{\NoviceTransitionMindWanderingToBreathFocus}{0.00}
\newcommand{\NoviceTransitionMindWanderingToMetaAwareness}{0.78}
\newcommand{\NoviceTransitionMindWanderingToRedirectAttention}{0.22}
\newcommand{\NoviceDwellMetaAwareness}{5.2}
\newcommand{\NoviceMetaAwarenessDMN}{0.703}
\newcommand{\NoviceMetaAwarenessVAN}{0.827}
\newcommand{\NoviceMetaAwarenessDAN}{0.419}
\newcommand{\NoviceMetaAwarenessFPN}{0.559}
\newcommand{\NoviceMetaAwarenessVIS}{0.354}
\newcommand{\NoviceMetaAwarenessSOM}{0.377}
\newcommand{\NoviceMetaAwarenessLIM}{0.631}
\newcommand{\NoviceTransitionMetaAwarenessToBreathFocus}{0.00}
\newcommand{\NoviceTransitionMetaAwarenessToMindWandering}{0.22}
\newcommand{\NoviceTransitionMetaAwarenessToRedirectAttention}{0.78}
\newcommand{\NoviceDwellRedirectAttention}{6.0}
\newcommand{\NoviceRedirectAttentionDMN}{0.411}
\newcommand{\NoviceRedirectAttentionVAN}{0.472}
\newcommand{\NoviceRedirectAttentionDAN}{0.773}
\newcommand{\NoviceRedirectAttentionFPN}{0.716}
\newcommand{\NoviceRedirectAttentionVIS}{0.330}
\newcommand{\NoviceRedirectAttentionSOM}{0.518}
\newcommand{\NoviceRedirectAttentionLIM}{0.313}
\newcommand{\NoviceTransitionRedirectAttentionToBreathFocus}{0.89}
\newcommand{\NoviceTransitionRedirectAttentionToMindWandering}{0.11}
\newcommand{\NoviceTransitionRedirectAttentionToMetaAwareness}{0.00}
\newcommand{\NoviceReturnEffort}{0.016}
\newcommand{\NoviceLearnedHabitReturn}{0.838}
\newcommand{\NoviceReturnEffortFitFirst}{0.025}
\newcommand{\NoviceReturnEffortFitLast}{0.017}
\newcommand{\NovicePracticeSteps}{12,000}
\newcommand{\NoviceAccessEvents}{167}
\newcommand{\NoviceBreathAccessPercent}{43.5}
\newcommand{\NoviceBreathRunSeconds}{3.4}
\newcommand{\NoviceDistractorAccessMWPercent}{80.5}
\newcommand{\NoviceClarityBF}{0.73}
\newcommand{\NoviceClaritySession}{0.58}
\newcommand{\NoviceBreathAccessPrior}{0.49}
\newcommand{\NoviceObservationPrecision}{2045}
\newcommand{\NoviceRestingClarity}{0.49}
\newcommand{\NoviceRegimeBeliefPercent}{100.0}

%% file: figures/robustness_macros.tex
\def\RobustnessModelVersion{clarity-v48}
\newcommand{\FiveRunExpertMetaAwarenessDetections}{36}
\newcommand{\FiveRunExpertMetaAwarenessDetectionsDistractor}{29}
\newcommand{\FiveRunNoviceMetaAwarenessDetections}{37}
\newcommand{\FiveRunNoviceMetaAwarenessDetectionsDistractor}{21}
\newcommand{\FiveRunExpertChecks}{28}
\newcommand{\FiveRunExpertEstablishedMWOnsets}{15}
\newcommand{\FiveRunExpertEstablishedMWOnsetsPerMinute}{1.9}
\newcommand{\FiveRunExpertEstablishedMWOnsetRuns}{5/5}
\newcommand{\FiveRunExpertEstablishedMWDistractorShare}{0.73}
\newcommand{\FiveRunExpertEstablishedMWEpisodes}{32}
\newcommand{\FiveRunExpertEstablishedMWNoOnsetShare}{0.59}
\newcommand{\FiveRunExpertEstablishedMWLongestNinety}{19.2}
\newcommand{\FiveRunNoviceChecks}{46}
\newcommand{\FiveRunNoviceEstablishedMWOnsets}{30}
\newcommand{\FiveRunNoviceEstablishedMWOnsetsPerMinute}{2.3}
\newcommand{\FiveRunNoviceEstablishedMWOnsetRuns}{5/5}
\newcommand{\FiveRunNoviceEstablishedMWDistractorShare}{0.53}
\newcommand{\FiveRunNoviceEstablishedMWEpisodes}{32}
\newcommand{\FiveRunNoviceEstablishedMWNoOnsetShare}{0.44}
\newcommand{\FiveRunNoviceEstablishedMWLongestNinety}{31.4}
\newcommand{\FiveRunExpertBreathRunUp}{5/5}
\newcommand{\FiveRunExpertBreathRunDown}{0/5}
\newcommand{\FiveRunExpertBreathAccessUp}{5/5}
\newcommand{\FiveRunExpertBreathAccessDown}{0/5}
\newcommand{\FiveRunExpertHabitReturnUp}{5/5}
\newcommand{\FiveRunExpertHabitReturnDown}{0/5}
\newcommand{\FiveRunExpertReturnEffortUp}{0/5}
\newcommand{\FiveRunExpertReturnEffortDown}{5/5}
\newcommand{\FiveRunNoviceBreathRunUp}{2/5}
\newcommand{\FiveRunNoviceBreathRunDown}{3/5}
\newcommand{\FiveRunNoviceBreathAccessUp}{5/5}
\newcommand{\FiveRunNoviceBreathAccessDown}{0/5}
\newcommand{\FiveRunNoviceHabitReturnUp}{5/5}
\newcommand{\FiveRunNoviceHabitReturnDown}{0/5}
\newcommand{\FiveRunNoviceReturnEffortUp}{0/5}
\newcommand{\FiveRunNoviceReturnEffortDown}{5/5}
\newcommand{\FiveRunExpertReentryOffsetEligible}{35}
\newcommand{\FiveRunExpertReentryOffsetTimed}{35}
\newcommand{\FiveRunExpertReentryOffsetMedian}{3.2}
\newcommand{\FiveRunExpertReentryOffsetRange}{1.9--3.5}
\newcommand{\FiveRunExpertReentryOffsetRecurred}{5}
\newcommand{\FiveRunNoviceReentryOffsetEligible}{34}
\newcommand{\FiveRunNoviceReentryOffsetTimed}{33}
\newcommand{\FiveRunNoviceReentryOffsetMedian}{4.6}
\newcommand{\FiveRunNoviceReentryOffsetRange}{3.8--5.2}
\newcommand{\FiveRunNoviceReentryOffsetRecurred}{11}

%% file: figures/fig1-tikz.tex
\begingroup
\resizebox{0.9\textwidth}{!}{%
\begin{tikzpicture}[x=1cm,y=1cm,font=\sffamily\small,
  regime/.style={rounded corners=6pt,line width=1pt,draw=black!60,fill=black!4,
    align=center,inner sep=6pt,text width=5.4cm,minimum height=2cm},
  flow/.style={draw=black!60,line width=1.6pt,-{Stealth[length=3.2mm]}},
  trans/.style={font=\sffamily\small,text=black!75}
]
\node[regime] (bf) at (0,3)      {\textbf{Breath Focus (BF)}\\[3pt]{\footnotesize\itshape sustained attention on the breath}};
\node[regime] (mw) at (5.5,6.2)  {\textbf{Mind Wandering (MW)}\\[3pt]{\footnotesize\itshape attention involuntarily drifts}};
\node[regime] (ma) at (11,3)     {\textbf{Meta Awareness (MA)}\\[3pt]{\footnotesize\itshape recognition of the lapse develops}};
\node[regime] (ra) at (5.5,-.2)  {\textbf{Redirect Attention (RA)}\\[3pt]{\footnotesize\itshape attention returns to the breath}};
\draw[flow] (bf.north) to[bend left=30] (mw.west);
\draw[flow] (mw.east)  to[bend left=30] (ma.north);
\draw[flow] (ma.south) to[bend left=30] (ra.east);
\draw[flow] (ra.west)  to[bend left=30] (bf.south);
\node[trans,anchor=east,align=right] at (0,5.7)   {Attention\\drifts};
\node[trans,anchor=west,align=left]  at (11,5.7)  {Recognition\\begins};
\node[trans,anchor=west,align=left]  at (11,.2)   {Disengage\\from distraction};
\node[trans,anchor=east,align=right] at (0,.2)    {Return focus\\to the breath};
\end{tikzpicture}}
\endgroup

%% file: figures/fig2-tikz.tex
\begingroup
\definecolor{conceptNeural}{HTML}{18365B}
\definecolor{conceptLatent}{HTML}{426B35}
\definecolor{conceptMeta}{HTML}{CB682A}
\definecolor{conceptUp}{HTML}{B43B37}
\definecolor{conceptDown}{HTML}{1976A3}
\definecolor{conceptBlanket}{HTML}{806184}
\resizebox{0.93\linewidth}{!}{%
\begin{tikzpicture}[x=1cm,y=1cm,font=\sffamily\small,
  shell/.style={rounded corners=12pt,line width=1.4pt},
  box/.style={rounded corners=5pt,line width=.9pt,fill=white,
    align=center,inner sep=7pt},
  sb/.style={box,draw=conceptMeta,inner sep=3pt,minimum width=4cm,minimum height=1cm},
  up/.style={draw=conceptUp,line width=1.2pt,-{Stealth[length=2.5mm]}},
  down/.style={draw=conceptDown,line width=1.2pt,-{Stealth[length=2.5mm]}},
  flow/.style={draw=black!55,line width=1pt,-{Stealth[length=2mm]}},  exchange/.style={draw=conceptBlanket,line width=1.1pt,-{Stealth[length=2.5mm]}},
  heading/.style={align=center,font=\sffamily\bfseries\normalsize},
  note/.style={font=\sffamily\footnotesize},
  tag/.style={font=\sffamily\footnotesize\itshape,align=center},
  lbl/.style={align=center,font=\sffamily\bfseries\small}
]
\draw[shell,draw=conceptNeural,fill=conceptNeural!3] (0,0)     rectangle (13.2,16.6);
\draw[shell,draw=conceptLatent,fill=conceptLatent!5] (.5,4.1)  rectangle (12.7,16.1);
\draw[shell,draw=conceptMeta,fill=conceptMeta!6]     (1,9.8)   rectangle (12.2,15.6);

\node[heading,text=conceptMeta] at (6.6,15.0) {Layer 3: Metacognition};

\node[box,draw=conceptMeta,minimum width=3.99cm,minimum height=3.6cm] (workspace) at (3.35,12.3) {};
\begin{scope}[shift={(3.35,12.3)}]
  \node[lbl,text=black] at (0,1.15) {Global workspace};
  \node[tag] at (0,.35) {content-specific};
  \node[note] at (0,-.45) {Accessed thoughtseed};
  \fill[conceptMeta] (0,-1.4) circle (.16);
\end{scope}

\node[sb] (mon) at (9.9,13.6) {\textbf{Monitoring clarity}\\[1pt]{\footnotesize\itshape graded}};
\node[sb] (reg) at (9.9,11.0) {\textbf{Return policy}\\[1pt]{\footnotesize\itshape exit timing and return}};
\node[sb,minimum width=3.4cm,fill=conceptMeta!10] (det) at (9.35,12.3)
  {\textbf{Lapse belief and detection}\\[1pt]{\footnotesize\itshape belief, event, known lapse}};
\draw[flow] (11.5,13.1) -- (11.5,11.5);
\draw[flow] (9.35,13.1) -- (det.north);
\draw[flow] (det.south) -- (9.35,11.5);
\draw[flow] (workspace.east) -- (det.west)
  node[midway,above=1pt,note,align=center] {Accessed\\content};

\node[box,draw=conceptMeta,minimum width=1.9cm,minimum height=1cm,inner sep=3pt] (prior) at (6.75,11.0)
  {\textbf{Access prior}\\[1pt]{\footnotesize\itshape learned}};
\draw[flow] (prior.west) -- (prior.west -| workspace.east);

\draw[down] (7.9,13.4) -- ([yshift=1.1cm]workspace.east)
  node[midway,above=1pt,note] {Modulates access};
\draw[up] (8.6,7.6) |- (12.45,8.15) |- (mon.east);
\node[note,anchor=south,align=center] at (11.65,8.22) {Thoughtseed\\activity};
\node[tag,text=conceptMeta,anchor=west] at (3.95,10.12) {capacity-limited access};

\node[heading,text=conceptLatent,align=center] at (7.4,8.85) {Layer 2: Low-dimensional\\latent causes};
\draw[box,draw=conceptLatent,fill=white] (1.2,4.55) rectangle (8.8,7.6);
\node[heading,text=conceptLatent,font=\sffamily\bfseries\small] at (6,7.2) {Thoughtseed repertoire};
\node[tag,text=conceptLatent] at (5,4.33) {small active subset (cf.\ consciousness prior)};

\node[box,draw=conceptLatent,minimum width=3.3cm,minimum height=1.9cm] (activecontents) at (3.35,5.7) {};
\begin{scope}[shift={(3.35,5.7)}]
  \node[lbl,text=black] at (0,.3) {Context-active\\thoughtseeds};
  \fill[conceptLatent!85] (-1.15,-.55) circle (.15) (-.7,-.55) circle (.15) (-.25,-.55) circle (.15);
  \fill[conceptMeta] (.2,-.55) circle (.15);
  \fill[conceptLatent!85] (1.05,-.55) +(0,.19) -- +(.19,0) -- +(0,-.19) -- +(-.19,0) -- cycle;
\end{scope}

\node[box,dashed,draw=conceptLatent,minimum width=3cm,minimum height=1.9cm] (dormant) at (7,5.7) {};
\begin{scope}[shift={(7,5.7)}]
  \node[align=center] at (0,.3) {Dormant\\thoughtseeds};
  \draw[conceptLatent!70,dashed,line width=.7pt] (-.9,-.55) circle (.15) (-.3,-.55) circle (.15)
                                                 (.3,-.55) circle (.15) (.9,-.55) circle (.15);
\end{scope}

\node[box,draw=conceptLatent,minimum width=2.8cm,minimum height=2.4cm] (fwd) at (10.45,6.0) {};
\begin{scope}[shift={(10.45,6.0)}]
  \node[lbl,text=black] at (0,.65) {Forwarded\\control};
  \node[note] at (0,-.05) {accessed content};
  \node[note] at (0,-.4) {known lapse};
  \node[note] at (0,-.75) {return policy};
\end{scope}

\fill[white] (2.15,8) -- (3.05,9.6) -- (3.05,10.5) -- (3.65,10.5) -- (3.65,9.6) -- (4.55,8) -- cycle;
\draw[conceptMeta,line width=.9pt,rounded corners=1.5pt]
  (3.05,10.5) -- (3.05,9.6) -- (2.15,8) -- (4.55,8) -- (3.65,9.6) -- (3.65,10.5);
\fill[conceptLatent!85] (2.85,8.4) circle (.13) (3.35,8.4) circle (.13) (3.85,8.4) circle (.13);
\draw[up] (3.35,6.65) -- (3.35,7.95);
\node[anchor=west,note] at (4.8,7.95) {Candidate contents};
\draw[down,-] (10.6,10.5) -- (10.6,8.3);
\draw[down] (10.6,8.0) -- (10.6,7.2);

\node[heading,text=conceptNeural] at (6.6,3.5) {Layer 1: High-dimensional brain networks};
\node[text=conceptNeural] at (6.6,2.95) {Neuronal substrate};
\node[box,draw=conceptNeural,text width=3.6cm,minimum height=2cm] at (3.4,1.45)
  {\textbf{Network activation\\profiles}\\[6pt]seven large-scale\\networks};
\node[box,draw=conceptNeural,text width=2.3cm,minimum height=2cm] at (10.65,1.45)
  {\textbf{Regime transitions}\\[4pt]{\footnotesize dwell-time\\distributions and\\transition table}};
\node[box,dashed,draw=conceptNeural,text width=2.9cm,minimum height=2cm] at (7.45,1.45)
  {Superordinate ensembles\\of neuronal packets\\[4pt]{\footnotesize\itshape conceptual motivation}};
\draw[up] (.9,1.45) |- (1.2,6.1);
\draw[down] (fwd.east) -- (12.3,6.0) -- (12.3,1.45);
\node[heading,text=conceptNeural,font=\sffamily\bfseries\small] at (6.6,-.5)
  {Internal states of the agent};

\draw[shell,dashed,draw=conceptBlanket,fill=conceptBlanket!5] (13.6,0) rectangle (17,4.3);
\node[lbl,text width=3cm,text=conceptBlanket] at (15.3,3.45) {Markov blanket\\of the agent};
\node[box,dashed,draw=conceptBlanket,text width=2.3cm,minimum height=1cm] (sensory) at (15.3,2.2) {Sensory states};
\node[box,dashed,draw=conceptBlanket,text width=2.3cm,minimum height=1cm] (actions) at (15.3,.8) {Active states};
\node[box,dashed,draw=black!55,text width=2.1cm,minimum height=3.1cm] (environment) at (19.1,1.5)
  {\textbf{External}\\\textbf{states}\\[5pt](environment)};
\draw[exchange] (environment.west |- sensory.east) -- (sensory.east);
\draw[exchange] (sensory.west) -- (13.2,2.2);
\draw[exchange] (13.2,.8) -- (actions.west);
\draw[exchange] (actions.east) -- (environment.west |- actions.east);

\draw[up] (13.5,9.9) -- (14.4,9.9);
\node[anchor=west,note] at (14.55,9.9) {Ascending evidence (perception)};
\draw[down] (14.4,9.3) -- (13.5,9.3);
\node[anchor=west,note] at (14.55,9.3) {Descending control (action)};
\draw[box,draw=black!60,inner sep=0pt] (13.5,8.5) rectangle (14.4,8.9);
\node[anchor=west,note] at (14.55,8.7) {Simulated};
\draw[box,dashed,draw=black!60,inner sep=0pt] (13.5,7.9) rectangle (14.4,8.3);
\node[anchor=west,note] at (14.55,8.1) {Conceptual, not simulated};
\end{tikzpicture}%
}
\endgroup

%% file: figures/fig3-tikz.tex
\begingroup
\definecolor{conceptNeural}{HTML}{18365B}
\definecolor{conceptLatent}{HTML}{426B35}
\definecolor{conceptMeta}{HTML}{CB682A}
\definecolor{conceptUp}{HTML}{B43B37}
\definecolor{conceptDown}{HTML}{1976A3}
\definecolor{learnedFill}{HTML}{F6E7A8}
\resizebox{\linewidth}{!}{%
\begin{tikzpicture}[x=1cm,y=1cm,font=\sffamily\small,
  shell/.style={rounded corners=12pt,line width=1.4pt},
  box/.style={rounded corners=5pt,line width=.9pt,fill=white,align=center,inner sep=3pt},
  cfg/.style={fill=black!12},
  lrn/.style={fill=learnedFill},
  tagbox/.style={box,inner sep=1.5pt,font=\sffamily\footnotesize},
  up/.style={draw=conceptUp,line width=1.2pt,-{Stealth[length=2.5mm]}},
  down/.style={draw=conceptDown,line width=1.2pt,-{Stealth[length=2.5mm]}},
  flow/.style={draw=black!55,line width=1pt,-{Stealth[length=2mm]}},
  heading/.style={font=\sffamily\bfseries\normalsize,align=left,anchor=west},
  boxhead/.style={font=\sffamily\bfseries\small,align=center},
  note/.style={font=\sffamily\footnotesize},
  tag/.style={font=\sffamily\footnotesize\itshape},
  lab/.style={font=\sffamily\footnotesize,inner sep=1pt,fill=white,fill opacity=.85,text opacity=1},
  badge/.style={circle,fill=black!75,text=white,inner sep=0pt,minimum size=.55cm,
    font=\sffamily\bfseries\footnotesize}
]
\draw[shell,draw=conceptNeural,fill=conceptNeural!3] (0,0)    rectangle (22.4,3.4);
\draw[shell,draw=conceptLatent,fill=conceptLatent!5] (0,4.2)  rectangle (22.4,11.6);
\draw[shell,draw=conceptMeta,fill=conceptMeta!6]     (0,12.2) rectangle (22.4,16.6);
\node[heading,text=conceptMeta]   at (.4,16.15) {Layer 3: Metacognition};
\node[heading,text=conceptLatent] at (.4,4.65)  {Layer 2: Low-dimensional latent causes};
\node[heading,text=conceptNeural] at (.4,3.0)   {Layer 1: High-dimensional brain networks};

\node[box,draw=conceptNeural,minimum width=7.4cm,minimum height=2.1cm] (net) at (5.1,1.35)
  {\textbf{Network activity} $\mathbf{x}_t$\\[1pt]{\footnotesize\itshape characteristic activation profiles}};
\node[box,cfg,draw=conceptNeural,minimum width=2.6cm,minimum height=2.1cm] (att) at (10.75,1.35)
  {\textbf{Network}\\\textbf{dynamics}\\[1pt]$\boldsymbol{\mu}_x(s)$\\{\footnotesize\itshape configured}};
\node[box,draw=conceptNeural,minimum width=8.6cm,minimum height=2.1cm] (reg1) at (17.5,1.35) {};
\begin{scope}[shift={(17.5,1.35)}]
  \node[boxhead] at (0,.72) {Regime $s_t$};
  \node[tag] at (0,.30) {exit lowered by broadcast stabilization $S_t$};
  \node[tag] at (0,-.04) {in MA, exit weighted by $\Lambda_t$, successor from $P(\pi)$};
  \node[box,cfg,draw=conceptNeural,inner sep=2pt,font=\sffamily\footnotesize\itshape] at (0,-.68)
    {configured dwell-time distributions and transition table};
\end{scope}
\draw[flow] (reg1.west) -- (att.east);
\draw[flow] (att.west) -- (net.east);

\node[box,draw=conceptLatent,minimum width=3.0cm,minimum height=2.2cm] (rho) at (2.0,7.25) {};
\node[boxhead] at (2.0,7.95) {Regime belief};
\node at (2.0,7.5) {$\rho_t$};
\node[box,cfg,draw=conceptLatent,inner sep=2pt,font=\sffamily\footnotesize\itshape] at (2.0,6.75)
  {configured\\transition model};

\draw[box,draw=conceptLatent] (4.0,5.8) rectangle (11.6,9.9);
\node[boxhead] at (7.8,9.6) {Thoughtseed repertoire};
\node[note,align=center] at (7.8,8.95) {switching Kalman filter};
\node[box,draw=conceptLatent,minimum width=3.4cm,minimum height=2.1cm] (zs) at (6.0,7.25) {};
\node[boxhead] at (6.0,7.65) {Context-active\\thoughtseeds};
\node at (6.0,7.0) {$\mathbf{z}^{*}_t$};
\fill[conceptLatent!85] (5.2,6.62) circle (.11) (5.5,6.62) circle (.11) (5.8,6.62) circle (.11);
\fill[conceptMeta] (6.1,6.62) circle (.11);
\fill[conceptLatent!85] (6.75,6.62) +(0,.14) -- +(.14,0) -- +(0,-.14) -- +(-.14,0) -- cycle;
\node[tagbox,lrn,draw=conceptLatent] (dec) at (5.15,6.2) {decoder, $\Pi_x$};
\draw[flow] (dec.west) -- ++(-0.3,0) |- ([yshift=-5mm]rho.east);
\node[box,cfg,draw=conceptLatent,minimum width=2.6cm,minimum height=2.1cm] (pri) at (9.9,7.25)
  {\textbf{Content}\\\textbf{priors} $\boldsymbol{\mu}_z(s)$\\[1pt]{\footnotesize\itshape configured}};
\draw[flow] (pri.west) -- (zs.east);
\draw[flow] (rho.east) -- (zs.west);

\node[box,draw=conceptLatent,minimum width=4cm,minimum height=2.4cm] (ctl) at (19.6,7.6)
  {\textbf{Descending control}\\[1pt]{\footnotesize $o_t$, $D_t$, $P(\pi)$, $\Lambda_t$}};

\node[box,lrn,draw=conceptMeta,minimum width=2.1cm,minimum height=2.0cm] (beta) at (1.85,14.1)
  {\textbf{Access}\\\textbf{prior} $\boldsymbol{\beta}$\\[1pt]{\footnotesize\itshape learned}};
\node[box,draw=conceptMeta,minimum width=3.4cm,minimum height=2.6cm] (ws) at (5.2,14.1)
  {\textbf{Global workspace}\\[1pt]$o_t$\\{\footnotesize\itshape accessed thoughtseed}\\[14pt]};
\fill[conceptMeta] (5.2,13.25) circle (.14);
\draw[flow] (beta.east) -- (ws.west);
\node[box,draw=conceptMeta,minimum width=3.4cm,minimum height=2.6cm] (mon) at (10.2,14.1)
  {\textbf{Monitoring clarity}\\[1pt]$m_t$\\[1pt]{\footnotesize check interval $e_t$}\\[6pt]};
\node[tagbox,lrn,draw=conceptMeta] at (10.55,12.8) {access counts, $m_{\mathrm{rest}}$};
\node[box,fill=conceptMeta!10,draw=conceptMeta,minimum width=3.5cm,minimum height=2.6cm] (rec) at (14.9,14.1)
  {\textbf{Lapse belief}\\\textbf{and detection}\\[1pt]{\footnotesize belief $q_t$, evidence $R_t$}\\
   {\footnotesize indicator $D_t$, onset $\Delta_t$}};
\node[box,draw=conceptMeta,minimum width=3.7cm,minimum height=2.6cm] (pol) at (19.6,14.1)
  {\textbf{Return policy}\\[1pt]$P(\pi)$, $\Lambda_t$\\
   {\footnotesize precision $\gamma_t=D_t\,m_tq_t$}\\[8pt]};
\node[tagbox,lrn,draw=conceptMeta] at (20.3,12.8) {learned habit $E$};
\draw[down] ([yshift=-4mm]mon.west) -- ([yshift=-4mm]ws.east);
\node[lab,align=center] at (7.7,14.0) {$m_{t-1}$};
\draw[flow] (mon.east) -- (rec.west) node[lab,midway,above=2pt,align=center] {$m_{t-1}$\\{\scriptsize for $q_t$}}
  node[lab,midway,below=2pt,align=center] {$m_t$\\{\scriptsize for $D_t$}};
\draw[flow] (mon.north) -- ++(0,0.5) coordinate (policyClarityTurn)
  -- (policyClarityTurn -| pol.north) node[lab,midway,above=2pt] {$m_t$}
  -- (pol.north);
\draw[flow] (rec.east) -- (pol.west) node[lab,midway,above=2pt] {$q_t$}
  node[lab,midway,below=2pt] {$D_t$};

\draw[up] (8.3,2.4) -- (8.3,5.8);
\node[lab,anchor=west] at (8.4,3.8) {$\mathbf{x}_t$};
\draw[up] (8.3,5.3) -| (rho.south);
\draw[up] (6.4,8.3) -- (6.4,12.8);
\node[lab,anchor=east] at (6.3,10.6) {$\mathbf{z}^{*}_t$};
\draw[up] (6.4,11.9) -| (8.9,12.8);
\fill[conceptUp] (6.4,11.9) circle (.07);
\node[lab,anchor=west] at (9.0,11.9) {$\mathbf{z}^{*}_t$};

\draw[flow] (10.9,8.3) -- (10.9,12.2);
\node[lab,anchor=east,align=right] at (10.8,10.75) {{\itshape derived from}\\{\itshape the priors, fixed}};

\draw[down] (6.75,12.8) |- (12.4,12.45) |- (ctl.west);
\node[lab] at (7.8,12.45) {$o_t$};
\draw[flow] (12.4,12.45) -| (13.75,12.8);
\fill[conceptDown] (12.4,12.45) circle (.07);
\draw[down] (rec.south) |- (18.4,10.6) -- (18.4,8.8);
\node[lab,anchor=south] at (16.6,10.6) {$D_t$};
\draw[down] ([xshift=-6mm]pol.south) -- ([xshift=-6mm]pol.south |- ctl.north);
\node[lab,anchor=west] at (19.15,10.8) {$P(\pi)$, $\Lambda_t$};
\draw[down] (ctl.south) -- (ctl.south |- reg1.north);
\node[lab,anchor=east,align=right] at (19.5,3.8) {applied at\\the next step};

\node[badge] at (net.north west) {1};
\node[badge] at (reg1.north west) {1};
\node[badge] at (rho.north west) {2};
\node[badge] at (zs.north west) {2};
\node[badge] at (ws.north west) {3};
\node[badge] at (mon.north west) {4};
\node[badge] at (rec.north west) {4};
\node[badge] at (pol.north west) {4};
\node[badge] at (ctl.north west) {5};

\node[box,cfg,minimum width=.8cm,minimum height=.5cm,inner sep=0pt,draw=black!60] at (.8,-1.0) {};
\node[anchor=west,note] at (1.4,-1.0) {Configured};
\node[box,lrn,minimum width=.8cm,minimum height=.5cm,inner sep=0pt,draw=black!60] at (4.6,-1.0) {};
\node[anchor=west,note] at (5.2,-1.0) {Learned};
\node[box,minimum width=.8cm,minimum height=.5cm,inner sep=0pt,draw=black!60] at (8.0,-1.0) {};
\node[anchor=west,note] at (8.6,-1.0) {Computed during the simulation};
\draw[flow] (14.4,-1.0) -- (15.3,-1.0);
\node[anchor=west,note] at (15.5,-1.0) {Within layer or configured};
\draw[up] (.4,-1.8) -- (1.3,-1.8);
\node[anchor=west,note] at (1.5,-1.8) {Ascending evidence (perception)};
\draw[down] (7.6,-1.8) -- (8.5,-1.8);
\node[anchor=west,note] at (8.7,-1.8) {Descending control (action)};
\node[badge] at (20.0,-1.0) {n};
\node[anchor=west,note] at (20.4,-1.0) {Update order};
\end{tikzpicture}%
}
\endgroup

%% file: figures/event_measures.tex
\begin{table}[tbp]
\centering
\caption{\textbf{Recognition and return across the five runs.} Each cell is the median of five within-run summaries, one per training seed 100--104 over the evaluation window, with the range across runs in parentheses. The proportion is a median of per-run proportions, not a pooled proportion. The two latency rows both start at a detection event and measure return at different levels, breath re-entry into the workspace and RA entry. They are conditional on different eligible cases, named in each row. A summary without eligible cases is unavailable. Definitions are in Section~\ref{sec:recognition}, eligibility in Supplementary Section~S3, paired differences in Supplementary Table~S4 and event counts in Supplementary Table~S5.}
\label{tab:event_measures}
\small
\begin{tabular}{>{\raggedright\arraybackslash}p{0.40\textwidth} >{\raggedright\arraybackslash}p{0.18\textwidth} c c}
\hline
Measure & Within-run summary & Expert & Novice\\
\hline
Recognition onsets with breath re-entry before or at recognition offset & proportion of all onsets & 0.38 (0.31--0.56) & 0.35 (0.25--0.44)\\
Recognition-to-breath latency (s), among those re-entries & median & 1.5 (1.2--2.2) & 2.4 (1.0--2.7)\\
Recognition-to-RA latency (s), qualifying MA episodes with a direct RA exit & median & 1.8 (0.5--2.3) & 2.8 (2.3--5.9)\\
Same-distractor recognition overlap (s), onsets with a distractor in the workspace & median & 0.6 (0.4--1.0) & 0.6 (0.5--0.8)\\
\hline
\end{tabular}
\end{table}

%% file: figures/robustness.tex
\def\RobustnessModelVersion{clarity-v48}
\begin{table}[!htbp]
\centering
\caption{\textbf{Five-run qualitative robustness check.} Paired expert-minus-novice differences over the final \StatsPlotSteps{} steps, given as the median (minimum--maximum) across training seeds 100--104. Positive dwell differences indicate longer expert mean dwell durations. Direction counts runs whose difference has the same sign as the median. When the median is zero, it counts zero differences. The latency and overlap rows compare within-run medians over each run's eligible events, with eligibility as in Table~2 of the main text and \S S3.}
\small
\begin{tabular}{p{0.56\textwidth} c c}
\hline
Measure & Paired difference & Direction\\
\hline
\multicolumn{3}{l}{\textit{Access, monitoring and regime dynamics}}\\
Breath Focus dwell (s) & $+6.3$ ($-0.1$--$+8.5$) & 4/5\\
Mind Wandering dwell (s) & $-10.2$ ($-14.4$--$-2.4$) & 5/5\\
RA$\rightarrow$BF transition & $-0.089$ ($-0.214$--$+0.400$) & 4/5\\
MW$\rightarrow$MA transition & $+0.000$ ($-0.167$--$+0.286$) & 3/5\\
Breath access run (s) & $+1.65$ ($+0.27$--$+1.83$) & 5/5\\
Mean $m_t$ over the window & $+0.139$ ($+0.103$--$+0.179$) & 5/5\\
Mean $m_t$ in Breath Focus & $+0.076$ ($+0.055$--$+0.093$) & 5/5\\
\multicolumn{3}{l}{\textit{Recognition and return}}\\
Recognition-to-RA latency, within-run median (s) & $-2.30$ ($-3.60$--$-0.60$) & 5/5\\
Return effort at MA exit (nats) & $-0.015$ ($-0.016$--$-0.014$) & 5/5\\
Onsets with breath re-entry & $+0.06$ ($-0.11$--$+0.28$) & 4/5\\
Recognition-to-breath latency, within-run median (s) & $-0.9$ ($-1.5$--$+0.5$) & 3/5\\
Same-distractor overlap, within-run median (s) & $-0.1$ ($-0.2$--$+0.4$) & 3/5\\
Distraction time with recognition & $-0.06$ ($-0.09$--$+0.03$) & 3/5\\
\quad in Mind Wandering & $-0.00$ ($-0.05$--$+0.04$) & 3/5\\
\quad in Meta Awareness & $-0.24$ ($-0.31$--$-0.06$) & 5/5\\
\hline
\end{tabular}
\label{tab:robustness}
\end{table}
\begin{table}[!htbp]
\centering
\caption{\textbf{Recognition onsets by outcome.} Final \StatsPlotSteps{} steps of training seeds 100--104, with counts summed over the five runs. Every recognition onset is counted by the regime at its step and its outcome in the definition of the breath re-entry proportion (Section~2.3 of the main text). The breath sensation was already in the workspace at onset, the breath re-entered before or at recognition offset, recognition ended first, or the recognition interval remained open at the window end without breath re-entry. The outcomes are exclusive and sum to the onsets $K$. Of the onsets that ended first, 70 ended through clarity, 1 through the belief and 0 through both. Re-entries over $K$ give a pooled proportion, which differs from the median of per-run proportions in Table~2 of the main text. The counts depend on the time spent in each regime and are not detection rates.}
\small
\begin{tabular}{l c c c c c}
\hline
Regime & Present & Re-entry & Ended first & Open, no re-entry & $K$\\
\hline
\multicolumn{6}{l}{\textit{Expert}}\\
BF & 1 & 0 & 0 & 0 & 1\\
MW & 2 & 5 & 31 & 0 & 38\\
MA & 7 & 29 & 0 & 0 & 36\\
RA & 6 & 1 & 1 & 0 & 8\\
all & 16 & 35 & 32 & 0 & 83\\
\hline
\multicolumn{6}{l}{\textit{Novice}}\\
BF & 1 & 0 & 0 & 0 & 1\\
MW & 12 & 8 & 38 & 0 & 58\\
MA & 13 & 23 & 1 & 0 & 37\\
RA & 2 & 3 & 0 & 0 & 5\\
all & 28 & 34 & 39 & 0 & 101\\
\hline
\end{tabular}
\label{tab:onset_outcomes}
\end{table}